\documentclass[twocolumn,aps,floatfix]{revtex4}
\usepackage{psfig}
\begin{document}
\title{Empirical formulas for the fermion spectra and Yukawa matrices}
\author{Javier Ferrandis} 
\email{ferrandis @ mac.com}
\homepage{http: // homepage.mac.com / ferrandis}
\affiliation{Department of Physics \& Astronomy \\
 University of Hawaii at Manoa \\
 2505 Correa Road \\
 Honolulu, HI, 96822}
\begin{abstract}
We present empirical relations that connect 
the dimensionless ratios of fermion masses for
the charged lepton, up-type quark and down-type quark sectors:
$ \left|V_{us}\right| \approx
\left[ \frac{m_{d}}{m_{s}} \right] ^{\frac{1}{2}}
\approx \left[ \frac{m_{u}}{m_{c}} \right]^{\frac{1}{4}} 
\approx 3 \left[ \frac{m_{e}}{m_{\mu}} \right]^{\frac{1}{2}}$ and 
$ \frac{1}{2} \left|\frac{V_{cb}}{V_{us}}\right| \approx \left[ \frac{m_{s}^{3}}{m_{b}^{2}m_{d}} \right] ^{\frac{1}{2}}
\approx \left[ \frac{m_{c}^{3}}{m_{t}^{2}m_{u}} \right] ^{\frac{1}{2}}
\approx  \frac{1}{9} \left[ \frac{m_{\mu}^{3}}{m_{\tau}^{2}m_{e}} \right]^{\frac{1}{2}}$. 
Explaining these relations from first principles imposes
strong constraints on the search for the theory of flavor. 
We present a simple set of normalized Yukawa matrices, 
with only two real parameters
and one complex phase, which accounts with precision for 
these mass relations and for the CKM matrix elements
and also suggests a simpler parametrization of the CKM matrix. 
The proposed Yukawa matrices accommodate
the measured CP-violation, giving a particular relation between standard model 
CP-violating phases, $\beta = {\rm Arg} \left[ 2 - e^{-i\gamma} \right]$. According to this relation
the measured value of $\beta$ is close to the maximum value that can be reached,
$\beta_{\rm max} =30^{\circ}$ for $\gamma=60^{\circ}$.
Finally, the particular mass relations between the quark and charged lepton sectors find their
simplest explanation in the context of grand unified models 
through the use of the Georgi-Jarlskog factor. 
\end{abstract}
\maketitle
\newpage
%
\section{Introduction}
Any theory of flavor must explain the fermion mass 
hierarchies as well as the quark mixing angles. 
Unfortunately, few patterns have been found 
in the measured values of fermion masses and mixing angles
that can guide us in the search for an underlying 
theory of flavor.
One of these, which has been known since 1968
\cite{Cabibborelation}, 
is the well known empirical relation between the down-quark mass, 
the strange-quark mass and the Cabibbo angle,
\begin{equation}
\left|V_{us} \right|\approx \left[ \frac{m_{d}}{m_{s}}\right] ^{\frac{1}{2}}.
\label{Cabibbo}
\end{equation}
This relation 
has driven the development of theories of flavor
over more than three decades,
starting in 1977 with the first attempt to explain it  
using family symmetries \cite{discrete,reviews}.
Another quark mass relation that has been known for some time is,
\begin{equation}
\left[ \frac{m_{d}}{m_{s}}\right] ^{\frac{1}{2}} \approx
\left[ \frac{m_{u}}{m_{c}}\right] ^{\frac{1}{4}}.
\label{druc}
\end{equation}
Inspired by these two relations, 
many of the theories of flavor proposed to date
have focused on generating Yukawa matrices that are 
polynomial in powers of $\lambda$, $\lambda \approx \left|V_{us}\right|$,
with coefficients of order 1 \cite{Froggatt:1978nt}. There is a third 
famous relation. It was argued as early as in 1979 
that at momenta larger than 
$10^{15}$~GeV quark and charged lepton masses are related by,
\begin{equation}
m_{d} = 3 m_{e}, \quad m_{\mu} = 3 m_{s},
\label{demus}
\end{equation}
An ingenious method was proposed to account for this relation by the use of 
SU(5) Clebsch-Gordan coefficients~\cite{Georgi:1979df}.
Other than these relations, it is usually claimed that the fermion masses follow 
scaling laws of the form $(m_{d},m_{s}) \simeq (\lambda^{4},\lambda^{2})m_{b}$
in the down-type quark sector, $(m_{u},m_{c}) \simeq (\lambda^{8},\lambda^{4})m_{t}$
in the up-type quark sector and $(m_{e},m_{\mu}) \simeq (\lambda^{4},\lambda^{2})m_{\tau}$
in the charged lepton sector.
As can be easily checked, however,
these scaling laws are qualitative 
and do not survive a precision analysis.

The measurement of the top quark mass in 1995 and
the continuous improvement in the extraction of 
other quark masses during the last decade motivate
a more systematic search for precise
empirical relations between dimensionless ratios of fermion masses in each fermion sector.
There are six independent fermion mass ratios of this kind, two for each fermion
sector. It is possible for hidden regularities 
to manifest themselves more clearly through higher order
dimensionless ratios of fermion masses, {\it i.e.} ratios of the form
$m^{2}_{a}/(m_{b}m_{c})$ or $m^{3}_{a}/(m^{2}_{b}m_{c})$. 
Indeed, as we show in this paper, there are some interesting
patterns underneath the measured values of the fermion masses. 
These new relations, which are not merely qualitative, are the following,
\begin{eqnarray}
\left|V_{us}\right| & \approx &
\left[ \frac{m_{d}}{m_{s}} \right] ^{\frac{1}{2}}
\approx \left[ \frac{m_{u}}{m_{c}} \right]^{\frac{1}{4}} 
\approx 3 \left[ \frac{m_{e}}{m_{\mu}} \right]^{\frac{1}{2}}, 
\\
\frac{1}{2} \left|\frac{V_{cb}}{V_{us}}\right| &\approx& \left[ \frac{m_{s}^{3}}{m_{b}^{2}m_{d}} \right] ^{\frac{1}{2}}
\approx \left[ \frac{m_{c}^{3}}{m_{t}^{2}m_{u}} \right] ^{\frac{1}{2}}
\approx  \frac{1}{9} \left[ \frac{m_{\mu}^{3}}{m_{\tau}^{2}m_{e}} \right]^{\frac{1}{2}}. 
\end{eqnarray}
We expect these two basic parameters, which we denote hereafter by
$\lambda \approx \left| V_{us}\right|$ and 
$\theta \approx \frac{1}{2} \left| V_{cb}\right| / 
 \left|V_{us}\right|$, 
to be connected with the fundamental parameters of the underlying 
theory of flavor. 

This paper is organized as follows. We begin in Sec.~\ref{correlations}
by systematically searching for correlations between dimensionless 
mass ratios in different fermion sectors up to order 3, 
{\it i.e.} up to ratios of the form $m^{3}_{a}/(m_{b}^{2}m_{c})$.
We review in the appendix the calculation of lepton and quark running masses
which are used in Sec.~\ref{correlations}.
In Sec.\ref{evolution} we analyze Yukawa renormalization corrections
that affect the studied mass relations when evolved with 
the renormalization scale, especially to the ratios including 
third generation fermion masses. 
In Sec.~\ref{hierarchy}
we show that, as a consequence of 
these new empirical formulas
the fermion mass hierarchies
can be expressed as a function of 
two basic parameters, $\lambda$ and $\theta$.
In Sec.~\ref{CKMcorrel} we show, neglecting CP-violation, 
that the absolute values of the CKM mixing matrix
elements can also be expressed as simple functions of the 
basic parameters $\lambda$ and $\theta$.
In Sec.~\ref{reconstr} we propose, neglecting CP-violation, 
a simple reconstruction of the quark Yukawa matrices
that accounts for the correlations found in the previous sections. 
In Sec.~\ref{CP} we introduce CP--violation in the textures proposed in 
Sec.~\ref{reconstr} and study its predictions for the CP-violating parameters.
In Sec.~\ref{CKMpredict} we study the precision predictions
for the lighter quark masses, CKM elements and charged lepton masses
arising from the texture proposed in Sec.~\ref{reconstr}.
In Sec.~\ref{leptonsec} we point out that the simplest solution to account for
the relations between the charged lepton sector and the quark sector
can be found in the the extension of the Standard Model
$SU(3)_{C} \times SU(2)_{L} \times U(1)_{Y}$  symmetry to the $SU(5)$
symmetry of Georgi and Glashow.
In Sec.~\ref{thetatheo} we speculate about the characteristics
of underlying flavor models that can reproduce these empirical mass relations. 
\section{Correlations between dimensionless fermion mass ratios \label{correlations}}
\begin{figure}[bht]    
\begin{minipage}{16.cm}
\psfig{figure=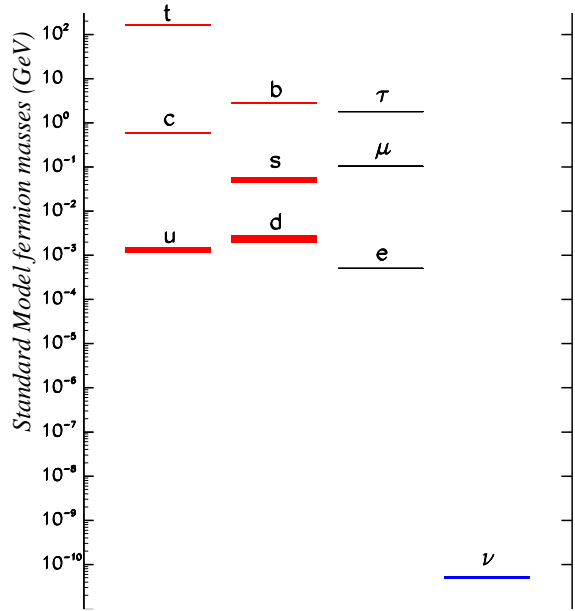,width=90mm}
\end{minipage}
\caption{\it  The fermion mass spectra.}
\label{fig:fermions}    
\end{figure}    
In this section we will look for patterns in the
dimensionless mass ratios of running fermion masses. 
Other than the fact that the first fermion generation is lighter than the second and
this is lighter than the third generation,
there are no other evident regularities in the fermion mass spectra,
as can be observed in Fig.~\ref{fig:fermions}.
Based on the experimental fact that the third generation is much heavier than the
first and second generations and that the quark mixing angles are small 
we hope that there is a simple
mechanism of flavor breaking which generates at some higher energy scale
a simple structure in the normalized Yukawa matrices. If this is the case it is plausible
that at such scale the normalized Yukawa matrices have the form,
\begin{equation}
\widehat{\cal Y} = 
\left[
\begin{array}{ccc}
0 & 0 & 0 \\
0 & 0 & 0 \\
0 & 0 & 1 \\
 \end{array}
\right] + {\cal O}( \lambda,\theta,\cdot \cdot \cdot), 
\label{Ynorm}
\end{equation}
where $\lambda,\theta,\cdot \cdot \cdot$ represent generically some perturbative flavor 
breaking parameters, {\it i.e.} $\lambda, \theta,\cdot \cdot \cdot  \ll 1$,
directly related to the underlying theory of flavor.
We note that in many flavor models proposed in the literature the flavor breaking 
is parametrized by a unique parameter $\lambda$. 
Therefore we expect the fermion
mass ratios in each one of the three fermion sectors: up-type quark, down-type quark
and charged lepton to be expressed as a simple polynomial functions of the flavor parameters:
$\lambda,\theta,\cdot \cdot \cdot$,
\begin{eqnarray}
\widehat{m}_{1} &=& \frac{m_{1}}{m_{3}} = f_{1}(\lambda,\theta,\cdot \cdot \cdot ), \\
\widehat{m}_{2} &=& \frac{m_{2}}{m_{3}} = f_{2}(\lambda, \theta, \cdot \cdot \cdot).
\end{eqnarray}
Let us assume to simplify the discussion that there are only two flavor breaking parameters:
$\lambda$ and $\theta$. In this case it would be possible to solve the previous system of equations
and obtain 
expressions for $\lambda$ and $\theta$ as a function of the fermion mass ratios,
\begin{eqnarray}
\lambda &=& \lambda ( \widehat{m}_{1}, \widehat{m}_{2} ),\\
\theta &=& \theta ( \widehat{m}_{1}, \widehat{m}_{2} ).
\end{eqnarray}
This can be done for each fermion sector separately. This makes plausible that 
underlying patterns manifest more clearly in higher order mass ratios, 
even though these can be
expressed as a function of the six basic fermion mass ratios.
When searching for mass relations between different fermion
sectors, it is convenient to calculate ratios of running fermion
masses at a common renormalization scale. If there are regularities
in the underlying Yukawa matrices, these will be manifested more clearly in the 
ratios of running fermion masses, not in the ratios of physical masses.
Using the running masses that
we have calculated in the appendix  
we obtain dimensionless mass ratios in the 
charged lepton, up--type quark and down--type
quark sectors. We calculate ratios of
order 1, $c^{[1]}$, order 2, $c^{[2]}$, and order 3, $c^{[3]}$.
These ratios are generically of the form, 
\begin{equation}
c_{ab}^{[1]} = \frac{m_{a}}{m_{b}}, \quad
c_{abc}^{[2]} = \frac{m_{a}^{2}}{m_{b}m_{c}}, \quad
c_{abc}^{[3]} = \frac{m_{a}^{3}}{m_{b}^{2} m_{c}},
\end{equation}
\begin{table*}
\begin{ruledtabular}
\begin{tabular}{|c||c|c||c|c||c|c|}
\hline
& \multicolumn{2}{c|}{\rm Charged leptons} & \multicolumn{2}{c|}{\rm Down-type quarks} &
\multicolumn{2}{c|}{\rm Up-type quarks}  \\
\hline 
\hline 
I&$m_{e}/m_{\mu}$   & $ (4.73711 \pm  0.00007 ) \times 10^{-3}$ &  
$m_{d} /m_{s}$ &  $(4.4 \pm 1.4)\times 10^{-2}$  &
$m_{u}/ m_c$ &  $(2.6 \pm 0.8)\times 10^{-3}  $\\ \hline 
II & $m_{\mu}/m_{\tau}$   & $  ( 5.882 \pm 0.001 ) \times 10^{-2}$ &  
$m_{s} /m_{b}$ &  $(2.4 \pm 0.4)\times 10^{-2}$  &
$m_{c}/ m_{t}$ &  $(3.7 \pm 0.6) \times 10^{-3} $\\ \hline 
III&$m_{e}m_{\mu}/m^{2}_{\tau}$   & $ (1.6390 \pm 0.0006 )\times 10^{-5} $ &  
$m_{d}m_{s} /m^{2}_{b}$ &  $(2.5 \pm 1.0)\times 10^{-5}$  &
$m_{u} m_{c}/ m^{2}_{t}$ &  $(3.65 \pm 1.5 )\times 10^{-8} $\\\hline 
IV &$m_{e}^{2}/m_{\mu} m_{\tau}$   & $(1.3199  \pm 0.0003 ) \times 10^{-6} $ &  
$m_{d}^{2} /m_{s} m_{b}$ &  $(4.7 \pm 2.5)\times 10^{-5}$  &
$m_{u}^{2}/ m_{c} m_{t}$ &  $(2.52 \pm 1.4)\times 10^{-8} $\\ \hline 
V &$m_{\mu}^{2}/m_{e} m_{\tau}$   & $ 12.417 \pm 0.002$ &  
$m_{s}^{2} /m_{d} m_{b}$ &  $0.53 \pm 0.26$  &
$m^{2}_{c}/m_{u} m_{t}$ &  $(1.45 \pm 0.71) $\\ \hline 
VI &$m_{e}^{3}/m^{2}_{\mu} m_{\tau}$   & $ (6.253 \pm 0.001) \times 10^{-9}$ &  
$m_{d}^{3} /m^{2}_{s} m_{b}$ &  $(2.1 \pm 1.8) \times 10^{-6}$  &
$m_{u}^{3}/ m^{2}_{c} m_{t}$ &  $(6.5 \pm 5.8) \times 10^{-11}$ \\ \hline 
VII &$m_{e}^{3}/m_{\mu} m^{2}_{\tau}$   & $ (3.678 \pm 0.001) \times 10^{-10}$ &  
$m_{d}^{3} /m_{s} m^{2}_{b}$ &  $(5.0 \pm 3.7)\times 10^{-8}$  &
$m_{u}^{3}/ m_{c} m^{2}_{t}$ &  $(2.44 \pm 2.0) \times 10^{-13}$ \\ \hline 
VIII & $m_{\mu}^{3}/m^{2}_{e} m_{\tau}$   & $2621.2 \pm 0.5$ &  
$m_{s}^{3} /m^{2}_{d} m_{b}$ &  $12 \pm 10$  &
$m_{c}^{3}/ m^{2}_{u} m_{t}$ &  $(560 \pm 455) $\\ \hline 
IX &$m_{e}m_{\mu}^{2}/m^{3}_{\tau}$   & $ (9.640 \pm 0.005) \times 10^{-7}$ &  
$m_{d}m_{s}^{2} /m^{3}_{b}$ &  $( 6.0 \pm 3.5) \times 10^{-7}$  &
$m_{u} m^{2}_{c}/ m^{3}_{t}$ &  $(1.4 \pm 0.8 )\times 10^{-10}$\\ \hline 
X &$m_{e}^{2}m_{\mu}/m^{3}_{\tau}$   & $(4.567 \pm 0.002) \times 10^{-9}$ &  
$m_{d}^{2} m_{s} /m^{3}_{b}$ &  $(2.7 \pm 1.6)\times 10^{-8}$  &
$m_{u}^{2} m_{c}/ m^{3}_{t}$ &  $(3.5 \pm 2.4)\times 10^{-13}$\\ \hline 
XI &$m_{\mu}^{3}/m_{e} m^{2}_{\tau}$   & $ 0.7304 \pm 0.0002 $ &  
$m_{s}^{3} /m_{d} m^{2}_{b}$ &  $(1.3 \pm 0.9) \times 10^{-2}$  &
$m_{c}^{3}/ m_{u} m^{2}_{t}$ &  $(5.4 \pm 3.5) \times 10^{-3}$ \\ 
\hline \hline
\end{tabular} 
\caption{\rm Dimensionless fermion mass ratios
in the charged lepton, up and down--type quark sectors 
calculated from measured values as explained in the text}
\label{ratios}    
\end{ruledtabular}
\end{table*} 
where $a,b,c=1,2,3$ are generation indices. 
Our numerical results are shown in table ~\ref{ratios}.
We have also included uncertainties for the mass ratios, $\Delta c$,
calculated using 
$\Delta c = \left| \partial c/\partial m_{a} \right| \Delta m_{a}$, where
$\Delta m_{a}$ are the uncertainties in the determination of 
running fermion masses.
The measured quark and charged lepton masses used as an input in our
calculations are explained in detail in the appendix.
We have compared analogous c coefficients
in the three different fermion sectors, looking for simple correlations of the form,
$c_{p} = r c_{q}$ or $c_{p} = c_{q}^{r}$ where $r$ is a low integer number
and $p,q=l,u,d$ denote similar ratios in the charged lepton, up-type or down-type
quark sectors.
 
 We have first searched for correlations between the mass ratios 
 in the up--type quark and down-type quark
sectors. 
We have only found two clear correlations. The first correlation appears for the
order one coefficient in entry I of table~\ref{ratios}.
The correlation appears between the ratios,
\begin{eqnarray}
\left[ \frac{m_{d}}{m_{s}}\right] ^{1/2} &=& 0.211 \pm 0.033, 
\label{Cabibbods} \\
\left[ \frac{m_{u}}{m_{c}}\right] ^{1/4} &=& 0.225 \pm 0.018,
\label{Cabibbouc}
 \end{eqnarray}
and the Cabibbo angle $\left| V_{us}\right|$.
These ratios have uncertainties
respectively of the order $\pm16$\% and $\pm8$\% of the central values.
It is convenient to show this correlation in an alternative form,
which makes it more manifest, 
\begin{eqnarray}
\left[\frac{m_{u}}{m_{c}}\right]^{1/4}  :
\left[\frac{m_{d}}{m_{s}}\right]^{1/2} &=&
1.06 \pm 0.25
\label{Cabibborelation}
\end{eqnarray}
This correlation has been known for some time. 
Curiously, we also find
an interesting correlation with the analogous ratio in the 
charged lepton sector,
\begin{eqnarray}
\left[ \frac{m_{d}}{m_{s}}\right] ^{1/2} :
\left[  \frac{m_{e}}{m_{\mu}} \right] ^{1/2}
&=& 3.06 \pm 0.48 
\label{Cabibbolepton} 
\end{eqnarray}
The $\pm 15 \%$ uncertainity in the calculation of this ratio comes from the
uncertainty in the determination of the lighter quark masses. This indicates that 
the following ratio in the charged lepton sector,
\begin{equation}
3 \left[  \frac{m_{e}}{m_{\mu}} \right] ^{1/2} = 0.20648 \pm 0.000002
\label{Cabibboemu}
\end{equation}
gives a numerical value very close to the Cabibbo angle and to the ratios 
in Eqs.~\ref{Cabibbods} and \ref{Cabibbouc}.
It was first pointed out in 1979 by Georgi and Jarlskog \cite{Georgi:1979df}
that at momenta larger than 
$10^{15}$~GeV quark and charged lepton masses seem to be related by,
\begin{equation}
m_{d} = 3 m_{e}, \quad m_{\mu} = 3 m_{s},
\label{demus}
\end{equation}
and that this relation could be explained by the use of 
SU(5) Clebsch-Gordan coefficients.
We want to emphasize that the previous correlations indicates that indeed
there is a very precise relation between the Cabibbo angle and the 
ratios of the fermion masses of the first and second generation, 
\begin{equation}
\left| V_{us} \right| \approx 
\left[ \frac{m_{d}}{m_{s}} \right] ^{1/2} \approx
\left[ \frac{m_{u}}{m_{c}} \right]^{1/4} \approx
3 \left[ \frac{m_{e}}{m_{\mu}} \right]^{1/2} 
\label{lambdaEq}
\end{equation}
We will see in Sec.~\ref{evolution} that,
as a very good approximation, this relation is renormalization
scale independent.
We have found only one more simple correlation amongst the 
dimensionless mass ratios shown in table~\ref{ratios}.
This appears for the
order three coefficient shown in the entry XI in table ~\ref{ratios}. It is convenient
to take the square root of the numbers shown in the table. 
In the up-type and down-type quark sectors we obtain,
\begin{eqnarray}
 \left[ \frac{m^{3}_{c}}{m_{t}^{2}m_{u}}\right]^{\frac{1}{2}} &=& 0.073 \pm 0.023 
 \label{thetaU}, \\ 
 \left[ \frac{m^{3}_{s}}{m_{b}^{2}m_{d}}\right]^{\frac{1}{2}} &=& 0.114 \pm 0.039.
\label{thetaD}
 \end{eqnarray}
Both ratios have an an important uncertainty, approximately $\pm 32$\%
of the central value, 
coming from the uncertainties in the extractions of the lighter quark masses.
It is convenient to quantify the correlation by taking the ratio,  
\begin{equation}
\left[ \frac{m^{3}_{s}}{m_{b}^{2}m_{d}}\right] ^{\frac{1}{2}} :
\left[ \frac{m^{3}_{c}}{m_{t}^{2}m_{u}}\right]^{\frac{1}{2}} = 1.5 \pm 1.0. 
\label{corthetaUD} 
\end{equation}
We note that there are two integer numbers, $\{1,2\}$, inside 
the error bars which could give us a simple correlation.
Furthermore, 
using in Eq.~\ref{thetaD} a recent lattice extraction of the 
strange quark mass mentioned in the appendix \cite{Aubin:2004ck}
instead of the sum rules extraction
the ratio in Eq.~\ref{corthetaUD} turns out to be closer to 1 
while the uncertainity is reduced,
\begin{equation}
\left[ \frac{(m^{\rm lat}_{s})^{3}}{m_{b}^{2}m_{d}}\right] ^{\frac{1}{2}} :
\left[ \frac{m^{3}_{c}}{m_{t}^{2}m_{u}}\right]^{\frac{1}{2}} = 0.8 \pm 0.5. 
\label{corthetaUDlat} 
\end{equation}
This lattice extraction of the strange quark mass 
has not been used throughout the main text because it has not yet 
been confirmed by other lattice QCD collaborations. 
If we compare with the analogous ratio between the down-type
sector and the charged lepton sector we obtain,
\begin{equation}
\left[ \frac{m_{\mu}^{3}}{m_{e}m_{\tau}^{2}} \right]^{\frac{1}{2}}  :
\left[ \frac{m^{3}_{s}}{m_{b}^{2}m_{d}}\right]^{\frac{1}{2}} = 
7.5 \pm 2.6 
\label{corthetaDL}
 \end{equation}
The uncertainty is again important, approx. $\pm34$\% of the central value, 
but it clearly points out that there are four integer numbers, $\{6,7,8,9\}$, inside 
the error bars which could give us a simple correlation. 
It is interesting to check the values predicted by multiplying 
by the inverse of these integer
factors the coefficient in the charged lepton sector. For instance, multiplying by $1/8$, $1/9$ and
$1/10$ we obtain the following values,
\begin{eqnarray}
\theta_{\tau}^{8}=
\frac{1}{8} \left[ \frac{m_{\mu}^{3}}{m_{e}m_{\tau}^{2}} \right]^{\frac{1}{2}} &=& 0.10681 \pm 0.00001, 
\label{thetaL8} \\
\theta_{\tau}^{9}=
\frac{1}{9} \left[ \frac{m_{\mu}^{3}}{m_{e}m_{\tau}^{2}} \right]^{\frac{1}{2}} &=& 0.09495 \pm 0.00001, 
\label{thetaL9} \\
\theta_{\tau}^{10}=
\frac{1}{10} \left[ \frac{m_{\mu}^{3}}{m_{e}m_{\tau}^{2}} \right]^{\frac{1}{2}} &=& 0.08545 \pm 0.00001. 
\label{thetaL10} 
\end{eqnarray}
If we compare these values with the values of the coefficients 
in Eqs.~\ref{thetaU} and \ref{thetaD} we observe that the integer 
factors 9 and 10 give us a number which is compatible with
the error bars of the coefficients in the up and down-type sector simultaneously.
We find specially interesting the appearance of the factor $9$ because as we will see 
in section~\ref{leptonsec} there is already a simple solution to explain this factor,
the Georgi-Jarlskog factor in GUT theories.
These results indicate that
there may be a second precise relation between the low energy
ratios of the fermion masses of the first, second and third generations,
\begin{equation}
\left[ \frac{m_{s}^{3}}{m_{b}^{2}m_{d}} \right] ^{1/2} \approx
\left[ \frac{m_{c}^{3}}{m_{t}^{2}m_{u}} \right] ^{1/2} \approx
\frac{1}{9} \left[ \frac{m_{\mu}^{3}}{m_{\tau}^{2}m_{e}} \right]^{1/2}.
\label{thetaEq} 
\end{equation}
The present uncertainities in the extraction of the lighter quark
masses do not allow us to determine if the relation works
at the $1\%$ level or just at the $40\%$ level. Surprisingly we must emphasize
that the exact empirical relation as given by Eq.~\ref{thetaEq} is inside the
$1\sigma$ experimental uncertainities for all of the three coefficients
as calculated in Eqs.~\ref{thetaU}, \ref{thetaD} and \ref{thetaL9}.
These can be observed more clearly in Fig.~\ref{fig:thetas}. 
We hope that near future improvements in the extraction of the ligther quark
masses, by the use of lattice QCD methods, could test with precision this
empirical formula. We note that there are only six independent mass ratios. We have already
found two simple and precise correlations linking the six of them. 
Therefore there cannot appear new correlations for other dimensionless 
mass ratios that cannot be expressed as a function of these two. 
\section{Fermion mass ratios and the Yukawa scale
\label{evolution}}
In Sec.~\ref{evolution} we searched for correlations between dimensionless
ratios of fermion masses at low energies. Nonetheless, it is known
that the top quark Yukawa coupling is of order 1,
since the top mass is of the same order than the electroweak scale.
This implies that Yukawa coupling corrections cannot
be ignored in the renormalization of the third generation fermion masses
to very high energies.
In models where the Higgs fields are flavor independent
approximate solutions that relate mass ratios at different renormalization
scales $\mu$ and $\mu_{0}$ are given by,
\begin{eqnarray}
\left(\frac{m_{d,s}}{m_{b}}\right)_{\mu} &\approx&  \left(\frac{m_{d,s}}{m_{b}}\right)_{\mu_{0}}  \xi_{b} , 
\label{bevol} \\ 
\left(\frac{m_{u,c}}{m_{t}}\right)_{\mu} &\approx&  \left(\frac{m_{u,c}}{m_{t}}\right)_{\mu_{0}} 
\xi_{t} ,  
\label{tevol} \\
\left(\frac{m_{e,\mu}}{m_{\tau}}\right)_{\mu} 
&\approx& \left( \frac{m_{e,\mu}}{m_{\tau}}\right)_{\mu_{0}} \xi_{\tau},
\label{tauevol} 
\end{eqnarray}
\begin{figure}[bht]    
\begin{minipage}{16.cm}
\psfig{figure=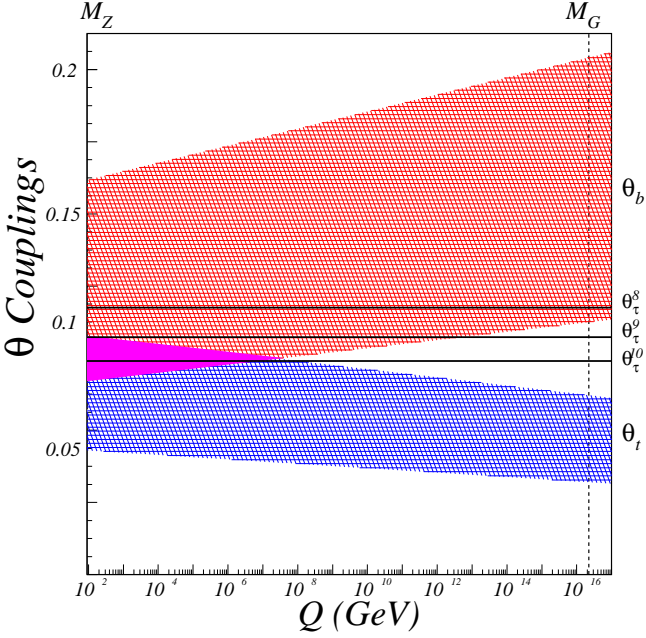,width=90mm}
\end{minipage}
\caption{\it  Renormalization scale evolution of the coefficients $\theta_{t}$,
$\theta_{b}$ and $\theta_{\tau}^{8,9,10}$ couplings 
and their uncertainities according to the SM RGE equations. 
The three coefficients and their uncertainities shown in the plot
were calculated using Eqs.~\ref{thetaU}, \ref{thetaD} and \ref{thetaL8}-\ref{thetaL10}. 
The ligther quark masses used were extracted using sum rules.}
\label{fig:thetas}    
\end{figure}    
In the case of the SM and the MSSM
these effects can be calculated approximately from available one-loop
renormalization group equations \cite{Babu:1992qn}. 
For the Standard Model we obtain 
$\xi_{b} = \xi_{t}^{-1} =\xi$ and $\xi$ is defined by, 
\begin{equation}
\xi \approx {\rm Exp} \left[  \frac{3}{32\pi^{2}} \ln \left( \frac{\mu}{\mu_{0}}\right) 
\left(1 - \left(\frac{m_{b}}{m_{t}}\right)^{2} \right) \right], 
\end{equation}
since $ m_{b} / m_{t}  \approx 2 \times 10^{-2}$ this is approximately,
\begin{equation}
\xi \approx \left( \frac{\mu}{\mu_{0}}\right) ^{ \frac{3}{32\pi^{2}}}.
\label{xism}
\end{equation}
Moreover the tau lepton Yukawa renormalization factor is very small,
\begin{equation}
\xi_{\tau} \approx  {\xi}^{\left(\frac{m_{\tau}}{m_{t}}\right)^{2}} 
\approx \left( \frac{\mu}{\mu_{0}}\right) ^{-\left(\frac{3}{32\pi^{2} 10^{4}}\right)}.
\end{equation}
We are interested in evaluating how the correlations found in Sec.~\ref{correlations}
evolve with the renormalization scale when including 
Yukawa corrections. To this end let us define
the dimensionless ratios,
\begin{equation}
\theta_{b} = \left[ \frac{m^{3}_{s}}{m_{b}^{2}m_{d}}\right] ^{\frac{1}{2}}, 
\quad
\theta_{t} = \left[ \frac{m^{3}_{c}}{m_{t}^{2}m_{u}}\right]^{\frac{1}{2}},
 \quad
\theta_{\tau} = \frac{1}{9} \left[ \frac{m^{3}_{\mu}}{m_{\tau}^{2}m_{e}}\right]^{\frac{1}{2}},
\end{equation}
Their evolution, using Eqs.~\ref{bevol}, \ref{tevol} and \ref{tauevol},
is given by,
\begin{equation}
\left(\frac{ \theta_{b} }{ \theta_{t}}\right)_{\mu} \approx 
\left(\frac{ \theta_{b} }{ \theta_{t}}\right)_{\mu_{0}} \xi^{2},
\quad
\left(\frac{ \theta_{b} }{ \theta_{\tau}}\right)_{\mu}  \approx 
\left(\frac{ \theta_{b} }{ \theta_{\tau}}\right)_{\mu_{0}}  \xi.
\end{equation}
If we assume that $\mu/\mu_{0} = M_{G}/M_{Z} \approx 
10^{14}$ we obtain $\xi \approx 1.36 $. If we extrapolate 
the mass relations up to the GUT scale using SM RGEs we obtain,
\begin{equation}
\left(\frac{ \theta_{b} }{ \theta_{t}}\right)_{M_{G}} \approx 2.8 \pm 1.0,
\end{equation}
and
\begin{equation}
\left(\frac{ \theta_{b} }{ \theta_{\tau}}\right)_{M_{G}} \approx 3.5 \pm 2.6,
\end{equation}
which must be compared with the low energy ratios calculated in Sec.~\ref{correlations}
\begin{equation}
\left(\frac{ \theta_{b} }{ \theta_{t}}\right)_{M_{Z}} = 1.5 \pm 1.0,
\end{equation}
and
\begin{equation}
\left(\frac{ \theta_{b} }{ \theta_{\tau}}\right)_{M_{Z}} = 7.5 \pm 2.6.
\end{equation}
Therefore the renormalization up to the GUT scale seems to
spoil the mass correlation between the up and down-type quark sector. 
These results are summarized in Fig.~\ref{fig:thetas} and
they may indicate that the Yukawa scale, the scale where the Yukawa couplings are generated,
is an intermediate scale much lower than the GUT scale or alternatively 
that is not correct to use SM RGEs 
in the evolution of the fermion masses up to the GUT scale. 
If we assume that the couplings evolve according to 
the MSSM RGEs the results depend on $\tan\beta$,
the ratio of Higgs expectation values in the MSSM. 
We obtain,
\begin{eqnarray}
\xi_{b} &\approx& \xi_{t}^{1/3} \approx 
{\rm Exp} \left[ - \frac{1}{16\pi^{2}} \ln \left( \frac{\mu}{\mu_{0}}\right)  \right],~ t_{\beta}\simeq 1, \\
\xi_{b} &\approx& \xi_{t} \approx 
{\rm Exp} \left[  - \frac{1}{4\pi^{2}} \ln \left( \frac{\mu}{\mu_{0}}\right)  \right],~ t_{\beta} \gg 1. 
\end{eqnarray}
Therefore we obtain the following scaling factors for low $\tan\beta$,
\begin{eqnarray}
\left(\frac{ \theta_{b} }{ \theta_{t}}\right)_{\mu} &\approx& 
\left(\frac{ \theta_{b} }{ \theta_{t}}\right)_{\mu_{0}} \xi^{4/3},\\
\left(\frac{ \theta_{b} }{ \theta_{\tau}}\right)_{\mu} &\approx& 
\left(\frac{ \theta_{b} }{ \theta_{\tau}}\right)_{\mu_{0}}\xi^{-2/3}. 
\end{eqnarray}
Here $\xi$ was defined in Eq.~\ref{xism} and using
$\mu/\mu_{0} = M_{G}/M_{Z}$ we
obtain $\xi^{4/3} \approx 1.5$ and 
$\xi^{-2/3} \approx 0.81$.
For large $t_{\beta}$ we obtain,
\begin{eqnarray}
\left(\frac{ \theta_{b} }{ \theta_{t}}\right)_{\mu} &\approx& 
\left(\frac{ \theta_{b} }{ \theta_{t}}\right)_{\mu_{0}}, \\
 \left(\frac{ \theta_{b} }{ \theta_{\tau}}\right)_{\mu} &\approx& 
\left(\frac{ \theta_{b} }{ \theta_{\tau}}\right)_{\mu_{0}}\xi^{-8/3}.
\end{eqnarray}
Here $\xi^{-8/3} \approx 0.44$ for
$\mu/\mu_{0} = M_{G}/M_{Z}$. Therefore in the MSSM, for both cases
low and large $\tan \beta$, we cannot extrapolate the mass relations
to scales as high as the GUT scale without spoiling the 
successful low energy mass relations. This again 
may be an indication that the Yukawa scale
is not so far from the electroweak scale.
We would like to point out that the ratios of first to second generation
fermion masses; $m_{d}/m_{s}$, $m_{u}/m_{c}$ and $m_{e}/m_{\mu}$,
receive tiny Yukawa renormalization factors. Therefore the mass relation,
\begin{equation}
\left[ \frac{m_{d}}{m_{s}} \right] ^{1/2} =
\left[ \frac{m_{u}}{m_{c}} \right]^{1/4} =
3 \left[ \frac{m_{e}}{m_{\mu}} \right]^{1/2}, 
\end{equation}
can be considered renormalization scale independent.
To sum up, this analysis indicates that the second empirical relation,
as given by Eq.~\ref{thetaEq}, may be optimal at some intermediate or low energy scale.
Nevertheless, the present uncertainities in the ligther quark masses 
are not small enough to allow us the determination of the scale at which 
this empirical formula is optimal.
\section{The fermion mass hierarchies \label{hierarchy}}
The empirical formulas found in the previous section
can be simply understood if the fermion mass hierarchies
are expressed as a function of two real parameters that we
will denote hereafter by $\theta$ and $\lambda$. Let us assume that 
the ratios of running masses can be written in the following form, 
\begin{eqnarray}
\left( m_{d}, m_{s} \right) &=& \left( \theta \lambda^{3}, \theta \lambda \right) m_{b}, 
\label{downhierarchy} \\
\left( m_{u}, m_{c} \right) &=& \left( \theta \lambda^{6}, \theta \lambda^{2} \right) m_{t}, 
\label{uphierarchy} \\
\left( m_{e}, m_{\mu} \right) &=& \left( \frac{1}{3} \theta \lambda^{3}, 3\, \theta \lambda \right) m_{\tau}, 
\label{lephierarchy}
\end{eqnarray}
We can easily prove that if this is the case we obtain immediately
the correct empirical mass relations, 
\begin{eqnarray}
\lambda & = &
\left[ \frac{m_{d}}{m_{s}} \right] ^{1/2} =
\left[ \frac{m_{u}}{m_{c}} \right]^{1/4} =
3 \left[ \frac{m_{e}}{m_{\mu}} \right]^{1/2} 
\label{lambdaEq2}
\\
\theta &=& \left[ \frac{m_{s}^{3}}{m_{b}^{2}m_{d}} \right] ^{1/2} =
\left[ \frac{m_{c}^{3}}{m_{t}^{2}m_{u}} \right] ^{1/2} =
\frac{1}{9} \left[ \frac{m_{\mu}^{3}}{m_{\tau}^{2}m_{e}} \right]^{1/2}.
\label{thetaEq2} 
\end{eqnarray}
In other words we can say the the hierarchies in Eqs.~\ref{downhierarchy}--\ref{lephierarchy}
solve the Eqs.~\ref{lambdaEq2}--\ref{thetaEq2}. We pointed out before that 
$\lambda$ corresponds approximately to the Cabibbo angle. 
The parameter $\theta$ may be considered a new flavor 
parameter that seems to suppress, in all three fermion sectors,
both first and second generation masses in the same 
amount with respect to the third generation.
The fact that $\theta$ and $\lambda$ connect 
different fermion sectors and that 
the correlations we have found work at a quantitative level
suggest that $\theta$ and $\lambda$ could be 
directly related to the underlying theory of flavor. 
We expect that any theory of flavor must be able to
explain these correlations from first principles,
and perhaps provide a prediction for $\theta$ and $\lambda$.

Although the important uncertainties in the current extraction of lighter quark masses
make impossible to know to what extent the 
relations in Eqs.~\ref{lambdaEq2} and \ref{thetaEq2} hold,
it is plausible to study the implications derived of
assuming that these relations are exact or almost exact.
If this were the case we are lead to assume that the 
charged lepton sector is providing us the 
most precise determination of 
the basic flavor parameters $\lambda$ and $\theta$,
$\lambda \approx 0.21$ and $\theta \approx 0.095$.

\section{The CKM mixing matrix hierarchies (neglecting CP-violation)\label{CKMcorrel} }
A theory of flavor must explain the fermion
mass hierarchies and the measured flavor mixing.
Therefore it is important for the reconstruction 
of the Yukawa matrices to study if it is possible to express the absolute values of 
the CKM matrix elements as simple functions of the two basic flavor parameters, 
$\theta$ and $\lambda$. We include for completeness a compilation
of the latest extractions of the elements of the 
CKM mixing matrix, $ \left| {\cal V}_{CKM}^{\hbox{exp}} \right| $,
\begin{equation}
\left[
 \begin{array}{ccc}
0.9739 \pm 0.0005 & 0.2224 \pm 0.0036 & 0.00357 \pm 0.00031 \\ 
 0.2224 \pm 0.0035  & 0.9740 \pm 0.0008 & 0.0415 \pm 0.0008 \\
\leq 0.005 & 0.0405 \pm 0.0035 & 0.99915 \pm 0.00015
\end{array}
\right].
\label{ckmexp}
\end{equation}
Here to obtain $\left|V_{ud}\right|$ two measurements, from superallowed Fermi transitions and
nuclear beta decay, have been combined, as in page 36 of 
the 2002 CERN Workshop on the CKM matrix ~\cite{Battaglia:2003in}. The value used for $\left|V_{us}\right|$ 
was calculated in page 37 of the same reference by requiring unitarity.
For $\left|V_{ub}\right|$ and $\left|V_{cb}\right|$ we use the latest extractions from B physics, as in 
page 6 of the same reference. For the rest of CKM elements we use 
the 2002 PDG compilation values \cite{Hagiwara:fs}. 
We fit each of the measured absolute values of the 
CKM matrix elements to simple functions 
of products of integer powers of the
fundamental parameters $\lambda$ and $\theta$.
We use the numerical values for $\theta$ and $\lambda$ as determined
from the charged lepton sector, {\it i.e.} $\theta \approx 0.095$ and $\lambda \approx 0.21$,
and look for correlations of the form $r ~\theta^{p}\lambda^{q}$,
where $p$ and $q$ are integer numbers and $r$ is a low integer or rational number.  
We obtain as the best fits to the measured CKM elements the functions,
\begin{eqnarray}
\left|V_{us}\right| &=& 0.2224 \pm 0.0036 ~ \approx  \lambda \\
\left|\frac{V_{cb}}{V_{us}} \right| &=& 0.187 \pm 0.006 \quad ~\approx 2 \theta \\
\left|\frac{V_{ub}}{V_{cb}} \right| &=& 0.086 \pm 0.009 \quad ~ \approx  \left( \frac{\lambda}{2} , \theta \right)
 \end{eqnarray}
We note that the ratio $\left| V_{ub}/V_{cb} \right|$ has an
important uncertainity which, in principle, would allow us to fit it
at $2\sigma$ to both terms $\theta$ or $\lambda /2$. 
The large experimental uncertainty in the entry $\left|V_{td}\right|$ 
does not allow us to implement a fit to $\theta$ and $\lambda$.
Therefore we obtain, ignoring CP-violating phases, 
the following structure for the CKM matrix as a function of the parameters
$\theta$ and $\lambda$,
\begin{equation}
{\cal V}_{\rm CKM} (\lambda, \theta) \approx
\left[
 \begin{array}{ccc}
1-\lambda^{2}/2 & -\lambda & a \\
\lambda & 1 - b^{2}  & -2 \theta \lambda \\
c & 2 \theta \lambda & 1 - 2 \theta^{2} \lambda^{2}
 \end{array}
\right], 
\end{equation}
where $a$, $b$ and $c$ must be considered unknown functions of $\theta$ and $\lambda$
which can be calculated requiring the matrix 
${\cal V}_{\rm CKM} (\lambda, \theta)$ to be unitary. This determines a system of
three equations which can be solved requiring unitarity to order ${\cal O}(\lambda^{3})$,
\begin{eqnarray}
V_{ud} V_{cd}+V_{us} V_{cs}+V_{ub} V_{cb} &=& {\cal O}(\lambda^{4}), \\
V_{ud} V_{td}+V_{us} V_{ts}+V_{ub} V_{tb} &=& {\cal O}(\lambda^{4}), \\
V_{td} V_{cd}+V_{ts} V_{cs}+V_{tb} V_{cb} &=& {\cal O}(\lambda^{4}).
\end{eqnarray}
We obtain $a=c=\theta \lambda^{2}$ and $b=\lambda^{2}/2+ 2 \theta^{2} \lambda^{2}$. Therefore
we obtain for $\left| {\cal V}_{\rm CKM} (\lambda, \theta) \right|$,
\begin{equation}
\left[
\begin{array}{ccc}
1-\lambda^{2}/2 &  \lambda & \theta \lambda^{2} \\
\lambda & 1-\lambda^{2}(1+4\theta^{2})/2  &  2 \theta \lambda \\
\theta \lambda^{2} & 2 \theta \lambda & 1 - 2 \theta^{2} \lambda^{2}
 \end{array}
\right], 
\label{ckmfit}
\end{equation}
We note that if we had chosen the correlation
$\left| V_{ub}/V_{cb} \right|  \approx  \theta$ instead of
$\left| V_{ub}/V_{cb} \right| \approx  \lambda/2$ the CKM matrix 
could not meet the unitarity requirement. 

We also note that one of the most interesting characteristics of the 
Yukawa matrices proposed in this section 
is that they account for the quark mass ratios and
CKM elements quantitatively with only two real parameters.
It is known that a general unitary matrix can be parametrized 
using three mixing angles and a complex phase. 
Our results indicate that only two real parameters seem to be necessary to 
account quite well for the absolute values of the CKM elements and quark mass
ratios. The CKM reconstructed matrix in Eq.~\ref{ckmfit} can be expressed,
using the standard PDG notation \cite{Hagiwara:fs} 
as the product of three rotation matrices, each around a different axis,
\begin{equation}
\left| {\cal V}_{\rm CKM}\right| \approx
R^{12}(\theta_{12})
R^{23}(\theta_{23})
R^{13}(\theta_{13})
\label{ckmrot}.
\end{equation}
where $\theta_{12} = \lambda$, $\theta_{23}= \theta \lambda$ and 
$\theta_{13}= \theta \lambda^{2}$.
Here $R^{12}(\theta_{12})$ is a rotation of angle $\lambda$
in the first-second generation plane,
\begin{equation}
R^{12}(\lambda) \approx
\left[
 \begin{array}{ccc}
1-\frac{\lambda^{2}}{2} & -\lambda & 0  \\
\lambda & 1-\frac{\lambda^{2}}{2}  & 0 \\
0 & 0 & 1 
\end{array}
\right],
\label{R12}
\end{equation}
and $R^{23}(\theta_{23})$  and $R^{13}(\theta_{13})$
are a rotations of angle $\theta \lambda$ and $\theta \lambda^{2}$ around
the second-third and first-third generation planes respectively,
\begin{equation}
R^{23}(2 \theta \lambda) \approx
\left[
 \begin{array}{ccc}
 1 & 0 & 0 \\
0 & 1 - 2 \theta^{2}\lambda^{2} & -2 \theta \lambda   \\
0 & 2\theta \lambda & 1- 2\theta^{2} \lambda^{2} 
\end{array}
\right],
\label{R23}
\end{equation}
and,
\begin{equation}
R^{13}(\theta \lambda^{2}) \approx
\left[
 \begin{array}{ccc}
1 - \frac{\theta^{2} \lambda^{4}}{2} & 0 & -\theta \lambda^{2} \\
0 & 1 & 0 \\
\theta \lambda^{2} & 0 & 1 - \frac{ \theta^{2}\lambda^{4}}{2} 
\end{array}
\right].
\label{R13}
\end{equation}
We have seen that present experimental data
allow us to express the fermion mass hierarchies and the absolute 
values of the CKM matrix elements 
as a function of two basic parameters $\theta$ and $\lambda$.
Throughout this section we have ignored the
presence of a CP-violating phase, which is required experimentally.
The presence of CP-violating phases could affect seriously the relations
between the absolute
values of some of the CKM elements and the parameters $\lambda$ and $\theta$
as proposed in this section. 
We will see in Sec.~\ref{CP} that  CP-violating phases can be included 
in the previous analysis giving an excellent fit to the data.
In the next section we will show that to leading order in $\lambda$ 
there is a unique set of Yukawa matrices that can reproduce 
these hierarchies. 
\section{Quark Yukawa matrices (neglecting CP-violation) \label{reconstr}}
In this section we  propose a
particular reconstruction of the quark Yukawa matrices that can explain 
the quark mass and mixing hierarchies found in previous sections. 
We restrict our discussion to symmetric mass matrices.
We will also assume that correct CKM elements or masses do not arise as approximate
cancellations requiring the tuning of different 
Yukawa matrix elements. 
It is convenient to define the $3\times 3$ normalized fermion mass matrices as,
$$
\widehat{\bf m}_{D}= \frac{1}{\widehat{m}_{b}} {\bf m}_{D},
\quad 
\widehat{\bf m}_{U} = \frac{1}{\widehat{m}_{t}} {\bf m}_{U}.
$$ 
Here ${\bf m}_{D,U}$ are the quark mass matrices and
$\widehat{m_{b}}$ and $\widehat{m_{t}}$
are normalized bottom and top quark masses,
which are defined as the ratio of bottom and top quark running masses 
over the largest eigenvalue of the respective normalized matrix.
We have the freedom to choose the entry 
(33) in the normalized mass matrices equal to 1,
which correspond to the heaviest eigenvalue. 
Although not diagonal in the gauge basis the matrix ${\bf m}_{D}$ 
can be brought to diagonal form in the mass basis by a biunitary diagonalization,
$({\cal V}^{d}_{L})^{\dagger} {\bf m}_{D} {\cal V}^{d}_{R}
=  \left( m_{d}, m_{s}, m_{b} \right)$.
Analogously the up-type quark mass matrix, ${\bf m}_{U}$,
can be brought to diagonal form in the mass basis by a biunitary diagonalization,
$ ({\cal V}^{u}_{L})^{\dagger} {\bf m}_{U} {\cal V}^{u}_{R}
=  \left( m_{u}, m_{c}, m_{t} \right)$.
The CKM mixing matrix is defined by
${\cal V}_{CKM} = {\cal V}^{u\dagger}_{L} {\cal V}^{d}_{L}$.

First we note that it is not possible to generate the Cabibbo angle,
$\lambda \approx \left| V_{us} \right|$, 
from the mixing between the first and second generations in the
up--type quark sector. The normalized charm quark mass is approximately $\theta \lambda^{2}$,
therefore to generate the correct magnitude for $\left| V_{us}\right|$ 
from $\left| {\cal V}^{u}_{L}\right|_{21}$, 
$\left| {\cal V}^{u}_{L}\right|_{21} \approx (\widehat{\bf m}_{U})_{12}/(\widehat{\bf m}_{U})_{22}$,
we would need $(\widehat{\bf m}_{U})_{12} \approx \theta \lambda^{3}$ but 
if it this were the case
the normalized up mass would be too heavy, $m_{u}/m_{t} \approx (\widehat{\bf m}_{U})_{12}^{2}/(\widehat{\bf m}_{U})_{22} \approx \theta \lambda^{4}$, which is in disagreement with 
the measured up-top mass hierarchy, 
$m_{u}/m_{t}\approx \theta \lambda^{6}$. 
Obtaining the correct up-quark mass give us a bound 
on the entries of the upper-left submatrix of the normalized 
up-type quark matrix, which must look like, 
\begin{equation}
(\widehat{\bf m}_{U})_{2 \times 2}^{\rm u-c} = 
\left[
 \begin{array}{ccc}
\leq  {\cal O}( \theta\lambda^{6}) & \leq {\cal O}(\theta \lambda^{4})  \\
\leq  {\cal O}( \theta \lambda^{4}) &  \theta \lambda^{2} 
\end{array}
\right]. 
\end{equation}
Therefore the Cabibbo angle must arise 
from first--second generation mixing in the down-type
quark sector. We pointed out in the previous sections that
the measured hierarchies in the down-type quark sector can be written as,
\begin{equation}
\frac{m_{d}}{m_{b}} \approx \theta \lambda^{3} << 
\frac{m_{s}}{m_{b}} \approx \theta \lambda < \left|V_{us}\right| \approx \lambda.
\end{equation}
We note that assuming $(\widehat{\bf m}_{D})_{12} = \theta \lambda^{2}$
and $(\widehat{\bf m}_{D})_{22} = \theta \lambda \approx m_{s}/m_{b}$, we obtain 
$(\widehat{\bf m}_{D})_{12}^{2}/(\widehat{\bf m}_{D})_{22} = \theta \lambda^{3} \approx m_{d}/m_{b}$
and $(\widehat{\bf m}_{D})_{12}/(\widehat{\bf m}_{D})_{22} = \lambda$.
This is consistent with a
down quark mass mainly generated from the first--second generation mixing as first pointed 
out by~\cite{discrete}, {\it i.e.},
\begin{equation}
\left(\widehat{\bf m}_{D}\right)_{2\times 2}^{d-s} = 
\left[
 \begin{array}{cc}
 0 & \theta \lambda^{2}  \\
  \theta \lambda^{2} &  \theta \lambda
\end{array}
\right],
\end{equation}
since in this case one correctly obtains 
$\left|{\cal V}_{L}^{d}\right|_{12} \approx
(\widehat{\bf m}_{D})_{12}/(\widehat{\bf m}_{D})_{22} = \lambda \approx \left| V_{us}\right|$.
Furthermore, as shown in the previous section, the measured hierarchies between
the CKM elements can be written as,
\begin{equation}
\left|V_{ub} \right| \approx \theta \lambda^{2} < 
 \left| V_{cb} \right| \approx 2 \theta \lambda 
 \simeq  (m_{s}/m_{b}) \approx \theta \lambda
 <  \left| V_{us} \right| \approx \lambda.
\end{equation}
We note that if we additionally assume 
that $(\widehat{\bf m}_{D})_{23} = 2 \theta \lambda$
then we obtain the correct $\left| V_{cb}\right|$, since
$\left| {\cal V}_{L}^{d}\right|_{23} = (\widehat{\bf m}_{D})_{23}/(\widehat{\bf m}_{D})_{33}=
2 \theta \lambda$. 
One can wonder if it would be possible to fully generate 
the CKM matrix entry $\left| V_{ub}\right|$
from the (23)-(12) mixing in the down-type quark sector, or in other words,
if we can assume that the normalized down-type quark matrix 
has a zero in the entry (13) and (31). 
A calculation of the diagonalization matrices show us that this possibility 
is not viable. It this were the case $\left|{\cal V}_{L}^{d}\right|$ would be given by,
 $\left|{\cal V}_{L}^{d}\right| \approx  \theta^{2} \lambda^{3} $ 
 which is two orders of magnitude below the measured value for 
 $\left|{\cal V}_{cb}\right|$,  $\left|{\cal V}_{cb}\right| \approx \theta \lambda^{2}$.
 Therefore we need to generate  $\left|{\cal V}_{cb}\right|$ directly from 
 a non-zero $(\widehat{\bf m}_{D})_{13}$ entry,
\begin{equation}
\widehat{\bf m}_{D} = 
\left[
 \begin{array}{ccc}
 0 & \theta \lambda^{2} & \theta \lambda^{2} \\
\theta \lambda^{2} &  \theta \lambda &  2 \theta  \lambda \\
 \theta \lambda^{2} & 2 \theta \lambda & 1
\end{array}
\right].
\label{mdrecons1}
\end{equation}
We must note here that  the possibility to fully generate $\left|V_{ub}\right|$ in the up-sector
is not viable, if that were the case we would generate a up-quark mass
one order of magnitude too heavy.  
This $\widehat{\bf m}_{D}$ matrix would predict successfully all the elements of the CKM mixing matrix,
assuming the mixing in the up-type sector does not affect the leading order predictions in $\lambda$.
This can be observed in the expressions for  the diagonalization matrix, 
${\cal V}_{L}^{d}$,  given by,
\begin{equation}
{\cal V}_{L}^{d} = 
\left[
 \begin{array}{ccc}
 1 - \frac{\lambda^{2}}{2}  &  \lambda &  -\theta \lambda^{2} \\
- \lambda  &   1 - \frac{\lambda^{2}}{2}(1 + 4 \theta^{2}) &  -2 \theta  \lambda \\
 -\theta \lambda^{2} & 2\theta \lambda & 1-  2 \theta^{2} \lambda^{2}
\end{array}
\right], 
\label{vdlvdr}
\end{equation}
Using this we obtain for the mass eigenvalues to leading order in $\lambda$,
\begin{equation}
({\cal V}^{d}_{L})^{\dagger} \widehat{\bf m}_{D} {\cal V}^{d}_{R}
\approx  \left(\theta\lambda^{3}, \theta \lambda, 1 +\theta^{2} \lambda^{2}\right).
\end{equation}
which is in perfect agreement with the reconstructed quark mass 
hierarchies in Eq.~\ref{downhierarchy}. In the previous reasoning
we assumed that the possible flavor mixing in the up-type quark sector 
does not affect to leading order the predictions for the CKM 
matrix which are generated in the down-type sector. If this were the case, 
there are two simple solutions which allow us to 
generate the correct up-quark mass:
directly from an entry (11),
\begin{equation}
\widehat{\bf m}_{U} = 
\left[
 \begin{array}{ccc}
 \theta \lambda^{6} & 0 & 0 \\
0 &  \theta \lambda^{2} &  0 \\
 0 & 0 &  1
\end{array}
\right], 
\label{mUrecons1a}
\end{equation}
 or from the first-second generation mixing,
\begin{equation}
\widehat{\bf m}_{U} = 
\left[
 \begin{array}{ccc}
 0 &  \theta \lambda^{4} & 0 \\
 \theta \lambda^{4} &  \theta \lambda^{2} &  0 \\
 0 & 0 &  1
\end{array}
\right],
\label{mUrecons1b}.
\end{equation}
Since both possibilities make the same predictions
for quark mass ratios and CKM elements both are
equivalent by a rotation of the quark fields.
It could be possible to generalize the previous solution 
to a solution that generates part of $\left| V_{cb} \right|$
from flavor mixing between second and third generations 
in the up-type quark sector. Let us assume that, 
\begin{equation}
\widehat{\bf m}_{D} = 
\left[
 \begin{array}{ccc}
 0 & \theta \lambda^{2} & \theta \lambda^{2} \\
\theta \lambda^{2} &  \theta \lambda &  (2-\epsilon) \theta  \lambda \\
 \theta \lambda^{2} & (2-\epsilon) \theta \lambda & 1
\end{array}
\right], 
\label{mDrecons2}
\end{equation}
and,
 \begin{equation}
\widehat{\bf m}_{U} = 
\left[
 \begin{array}{ccc}
 \theta \lambda^{6}   &  0 & 0 \\
 0 &  \theta \lambda^{2} &   - \epsilon \theta \lambda  \\
 0 &  - \epsilon \theta \lambda  &  1
\end{array}
\right].
\label{mUrecons2}
\end{equation}
In this case the 
diagonalization matrices are respectively,
\begin{equation}
{\cal V}_{L}^{d} = 
\left[
 \begin{array}{ccc}
 1 - \frac{\lambda^{2}}{2}  &  -\lambda &  -\theta \lambda^{2} \\
\lambda  &  1 - \frac{\lambda^{2}}{2}(1 + \eta^{2}\theta^{2}) &   \eta \theta  \lambda \\
(\epsilon -1 ) \theta \lambda^{2}  & \eta \theta \lambda & 1- \frac{1}{2} \eta^{2}\theta^{2} \lambda^{2} 
\end{array}
\right], 
\end{equation}
where $\eta= (2-\epsilon)$ and,
\begin{equation}
{\cal V}_{L}^{u} = 
\left[
 \begin{array}{ccc}
 1 & 0 & 0 \\
0  & 1 - \epsilon^{2} \frac{\theta^{2} \lambda^{2}}{2}  &  - \epsilon \theta \lambda  \\
0 &  \epsilon \theta \lambda  &   1 - \epsilon^{2} \frac{\theta^{2} \lambda^{2}}{2} 
\end{array}
\right].
\label{vulvurA}
\end{equation}
These two solutions; the one given by Eqs.~\ref{mdrecons1} and \ref{mUrecons1a}-\ref{mUrecons1b}
or the one given by Eqs.~\ref{mDrecons2} and \ref{mUrecons2}  
are indistinguishable in their predictions
for quark mass ratios and CKM elements to first order in powers of $\lambda$. 
Both reproduce the correct form for the CKM matrix in Eq.~\ref{ckmfit}.
Therefore they are equivalent and can be related by a rotation of the quark fields.
We note 
that from their diagonalization we obtain the correct empirical
expressions for $\lambda$ and $\theta$ 
as a function of dimensionless quark mass ratios
(see Eqs.~\ref{lambdaEq} \& \ref{thetaEq}). To first order,
\begin{eqnarray}
\lambda &\approx& \left( \frac{m_{d}}{m_{s}}\right)^{\frac{1}{2}} 
\approx \left( \frac{m_{u}}{m_{c}}\right)^{\frac{1}{4}}, \\
\theta &\approx& 
\left(  \frac{m_{s}^{3}}{m_{b}^{2} m_{d}} \right)^{\frac{1}{2}} \approx
\left(  \frac{m_{u}^{3}}{m_{c}^{2} m_{t}} \right)^{\frac{1}{2}},
\end{eqnarray}
Nevertheless, there are in principle other solutions
that are not equivalent to the family of solutions here proposed
and make the same predictions to leading order. These
could be differentiated in their precision predictions 
for mass ratios and CKM elements 
when including higher orders in powers of $\lambda$.
\section{Introducing CP-violation \label{CP}}
\begin{figure}[bht]    
\begin{minipage}{16.cm}
\psfig{figure=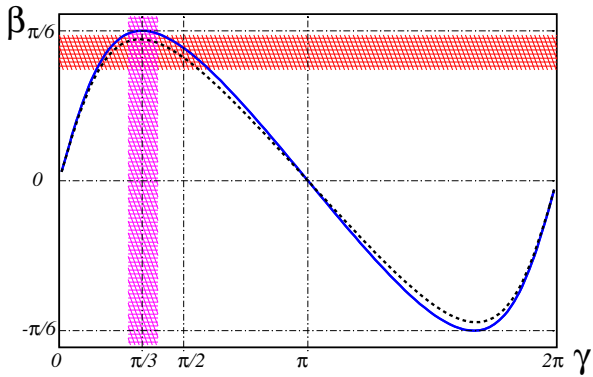,width=90mm}
\end{minipage}
\caption{\it  Relation between CP-violating phases $\beta$ and $\gamma$ as predicted
by the quark Yukawa matrices in Eqs.~\ref{mUrecons1a} and \ref{mdCPgamma}. The horizontal hatched strip 
corresponds to the measurement $\sin(2\beta)_{\rm exp}= 0.78 \pm 0.08$. The vertical hatched strip 
corresponds to the 1$\sigma$ global fit for the angle $\gamma$, 
$\gamma_{\rm fit} = 61^{\circ}\pm 11^{\circ}$. The solid curve corresponds to the 
leading order relation between $\beta$ and $\gamma$ 
given by Eq.~\ref{betagamma}. The dotted curved corresponds
to the next to leading order relation given by Eq.~\ref{betagamma2}}     
\label{fig:betagamma}    
\end{figure}    
We have seen in the previous section 
that a set of two parameter Yukawa 
matrices of the form given in Eqs.~\ref{mdrecons1} and \ref{mUrecons1a} 
represents a family of solutions, in the basis where the up-type Yukawa matrix is diagonal, 
that can account for the quark mass ratios and the absolute values 
of the CKM matrix elements.
It is possible to introduce complex
phases in this picture to account for the measured CP-violation 
without spoiling these successful predictions.
In order to do so we promote the real symmetric 
matrix in Eq.~\ref{mdrecons1} to be hermitian. In the most general hermitian 
case we can introduce complex phases in the form,
\begin{equation}
\widehat{\bf m}_{D} = 
\left[
 \begin{array}{ccc}
 0 & e^{i\psi_{1}} \theta \lambda^{2} &  e^{i\psi_{2}} \theta \lambda^{2} \\
 e^{-i\psi_{1}}\theta \lambda^{2} &  \theta \lambda &  2  e^{i\psi_{3}}\theta  \lambda \\
  e^{-i\psi_{2}} \theta \lambda^{2} & 2 e^{-i\psi_{3}}\theta \lambda & 1
\end{array}
\right].
\label{mdCP}
\end{equation}
By a redefinition of the phases of the quark fields this can be simplified to,
\begin{equation}
\widehat{\bf m}_{D} = 
\left[
 \begin{array}{ccc}
 0 &  \theta \lambda^{2} &   e^{-i\gamma} \theta \lambda^{2} \\
 \theta \lambda^{2} &  \theta \lambda &  2  \theta  \lambda \\
  e^{i\gamma} \theta \lambda^{2} & 2 \theta \lambda & 1
\end{array}
\right].
\label{mdCPgamma}
\end{equation}
where $\gamma = - (\psi_{2} -\psi_{1} - \psi_{3})$. 
If this were the case we obtain for the CKM matrix,
${\cal V}_{CKM} = {\cal V}_{L}^{d}$, to leading order in $\lambda$,
\begin{equation}
\left[
 \begin{array}{ccc}
 1 -  \frac{\lambda^{2}}{2}  & \lambda &  - e^{ -i \gamma}\theta \lambda^{2} \\
- \lambda  &  1 - \frac{\lambda^{2}}{2}(1  + 4 \theta^{2}) &   -2\theta  \lambda \\
(e^{i \gamma} -2 ) \theta \lambda^{2}  &  2\theta \lambda & 1-  2 \theta^{2} \lambda^{2} 
\end{array}
\right].
\label{CKMCP}
\end{equation}
The angle $\gamma$ introduced this way coincides with the 
standard definition, 
\begin{equation}
\gamma = {\rm Arg} \left[ - \frac{V_{ud}V^{*}_{ub}} {V_{cd}V^{*}_{cb}} \right],
\end{equation}
Furthermore the angle $\beta$ is given by,
\begin{equation}
\beta = {\rm Arg} \left[ - \frac{V_{cd}V^{*}_{cb}} {V_{td}V^{*}_{tb}} \right]. 
\end{equation}
The angle $\alpha$ can be obtained from the relation,
$\alpha + \beta + \gamma = 180^{\circ}$.
The hermitian matrix of the form given by
Eq.~\ref{mdCPgamma} predicts a simple 
relation between the angles $\beta$ and $\gamma$, which 
to leading order is, 
\begin{equation}
\beta =  {\rm Arg} \left[ 2 - e^{-i \gamma}\right]
\label{betagamma}
\end{equation}
The angle $\beta$ can been determined with $7\%$ accuracy
using the experimental determination of $\sin 2\beta$,  
$\sin 2\beta =0.78 \pm 0.08$. We obtain $\beta_{\rm exp} =25.7^{\circ}\pm 3.5^{\circ}$. 
This indicates that the value of $\beta$ that nature has chosen 
is close to the maximum value that $\beta$ can reach 
in the case under consideration,
$\beta_{\rm max}= 30^{\circ}$, which appears for $\gamma=60^{\circ}$,
\begin{equation}
\frac{ {\rm d}\beta}{{\rm d} \gamma} = 0 
\rightarrow \gamma=60^{\circ}\rightarrow \alpha=90^{\circ}.
\label{betamax}
\end{equation}
We note that this does not correspond with the case of 
maximal CP-violation. 
If we compute the determinant of Jarlskog matrix,
${\cal C} =  \left[\widehat{\bf m}_{U}\widehat{\bf m}_{U}^{\dagger},
\widehat{\bf m}_{D}\widehat{\bf m}_{D}^{\dagger}\right]$, we obtain,
\begin{equation}
{\rm det}~{\cal C}
=  2 {\cal J }\theta^{4} \lambda^{6} + {\cal O}(\lambda^{11}),
\end{equation}
where ${\cal J}$ is the Jarlskog parameter, the invariant measure of CP-violation,
which in our case is given by,
\begin{equation}
{\cal J} = 2 \sin(\gamma) \theta^{2} \lambda^{4}\left( 1 + {\cal O}(\theta \lambda)\right).
\label{Jarlskog}
\end{equation}
We note that the maximal CP-violation case corresponds to $\gamma=\pi/2$
and using the Eq.~\ref{betagamma} this corresponds to, 
\begin{equation}
\frac{ {\rm d}{\cal J}}{{\rm d} \gamma} = 0 
\rightarrow \gamma=90^{\circ}\rightarrow \beta=26.6^{\circ}.
\label{maxCP}
\end{equation}
Although there is a simple relation, see Eq.~\ref{betagamma}, between 
$\beta$ and $\gamma$ the angle $\gamma$,
as can be seen in figure~\ref{fig:betagamma}, 
cannot be determined with good precision from that 
relation and the experimentally determined
value of $\beta$. We obtain, $\gamma_{\rm theo} = 65^{\circ} \pm 38^{\circ}$,
which is in agreement with the 2004 winter global fit
of the CKM elements. obtained using the results of program 
CKMFitter \cite{Hocker:2001xe},
\begin{equation}
\gamma_{\rm exp}  =  61^{\circ} \pm 11 ^{\circ} \\
\end{equation}
Alternatively we can use this experimental value of $\gamma$ to 
predict $\beta$ from the Eq.~\ref{betagamma}, as can be seen 
on Fig.~\ref{fig:betagamma}. We obtain to leading order, 
\begin{equation}
\beta_{\rm theo}  =  29.4^{\circ} \pm 0.05 ^{\circ} \\
\end{equation}
which corresponds to $\sin(2\beta)_{\rm theo} = 0.855\pm 0.001$.
The Jarlskog parameter is determined experimentally to be 
${\cal J} = (3.0 \pm 0.3) \times 10^{-5}$. The use of the Jarlskog parameter
does not allow us to extract $\gamma$ with better precision
because of the uncertainties in the determination of $\lambda$ and $\theta$.
On the other hand our expression for ${\cal J}$ predicts an interesting relation
between ${\cal J}$ and the quark masses, 
\begin{equation}
{\cal J} \approx
2 \frac{  m_{d} m_{s}}{ m_{b}^{2}}
\sin (\gamma) \left( 1 + {\cal O}(\theta \lambda)\right).
\end{equation}
Finally, we note that there two non-trivial characteristics in the relation between
$\beta$ and $\gamma$, Eq.~\ref{betagamma}, as predicted by the CKM matrix 
in Eq.~\ref{CKMCP}: first there is no dependence
on $\lambda$ or $\theta$ to first order and second and most important
the relation agrees with the experimental measurements of $\beta$ and $\gamma$.
\section{Predictions for masses and mixings
\label{CKMpredict}}
In this section we want to show that the simple three parametric
Yukawa matrices proposed in
Sec.~\ref{CP} can fit all the experimental data on the fermion spectra with precision.
We study in this section precision predictions for the 
lighter quark masses and CKM elements. Let us assume that the $\widehat{\bf m}_{D}$
Yukawa matrix is given by Eq.~\ref{mdCPgamma}
while the $\widehat{\bf m}_{U}$ matrix is given by Eq.~\ref{mUrecons1a}.
If we postulate these up-- and down--type quark Yukawa matrices 
we can express the CKM elements and the 
quark mass ratios as a function of $\theta$, $\lambda$ and $\gamma$.  
The down-type quark mass ratios can be calculated to the next to leading
order in $\lambda$ from the diagonalization of Eq.\ref{mdCPgamma}, 
\begin{eqnarray}
\frac{m_{d}}{m_{s}} &=& \lambda^{2} ( 1 - \theta \lambda ( 4c_{\gamma} - 9) ), \\
\frac{m_{s}}{m_{b}} &=& \theta \lambda ( 1 - 4 \theta \lambda + \lambda^{2}  ). 
\end{eqnarray}
Here $c_{\gamma} = \cos (\gamma)$.
The up-type quark mass ratios are given by, 
\begin{equation}
\frac{m_{u}}{m_{c}} = \lambda^{4},\quad
\frac{m_{c}}{m_{t}} = \theta \lambda^{2} 
\label{usecmara}
\end{equation}
while the absolute values of 
CKM matrix elements to the 
next to leading order in $\lambda$ are given by,
\begin{eqnarray}
\left|V_{us}\right| &=&  \lambda - 2 (c_{\gamma} -2) \theta \lambda^{2} +{\cal O}(\lambda^{3}), \label{vus} \\
\left| V_{ud}\right| &=& 1 -  \frac{1}{2} \lambda^{2} +2 (c_{\gamma} -2) \theta \lambda^{3}  , \label{vud} \\
\left| V_{ub}\right| &=&  \theta \lambda^{2}  + 2 c_{\gamma} \theta^{2} \lambda^{3}, \\
\left| V_{cs}\right| &=& 1 -  \frac{1}{2} \lambda^{2}(1 +4\theta^{2}) + 
2 \theta \lambda^{3} ( c_{\gamma}  -2), \\
\left|V_{cb}\right| &=& 2 \theta \lambda ( 1 + \theta \lambda) ,\\
\left| V_{td}\right| &=& 
(5-4c_{\gamma})^{\frac{1}{2}} \left( \theta \lambda^{2}  + 4 \theta^{2} \lambda^{3}\right) , \\
\left|V_{ts}\right| &=&  2 \theta \lambda  + 2 \theta^{2} \lambda^{2} + (c_{\gamma}-1) \theta \lambda^{3}, \\
\left| V_{tb} \right| &=& 1 -  2 \lambda^{2}\theta^{2}- 4\theta^{3} \lambda^{3} .\label{vtb}
\end{eqnarray}
and $\left|V_{cd}\right| = \left|V_{us}\right|$. 
\begin{table}
\begin{tabular}{|c|c|}
\hline 
\hline 
\multicolumn{2}{|c|}{experimental input parameters } \\ 
\hline
$\left| V_{us}\right|$ &  { $0.2225 \pm 0.0035$} \\
$m_{c}/m_{t}$ &{ $(3.7 \pm 0.4)\times 10^{-3}$} \\
$\gamma$ & { $61^{\circ} \pm 11^{\circ}$} \\
$m_{t}^{\rm pole}$ &  { $174.3 \pm 5.1 ~{\rm GeV}$} \\
$m_{b}(m_{b})_{\overline{\rm MS}}$ & {$4.2 \pm 0.1 ~{\rm GeV}$} \\
$m_{\tau}^{\rm pole}$ & {$1.7769 \pm 0.0003 ~{\rm GeV}$} \\
\hline
\hline 
\multicolumn{2}{|c|}{predictions} \\ 
\hline
$\lambda$ &   {$0.211 \pm 0.007$}\\
$\theta$ &   $0.083 \pm 0.014$ \\
$\sin(2\beta)$ & { $0.824 \pm 0.004$} \\
$\left| V_{ud}\right|$ & {$0.975 \pm 0.002$} \\
$\left| V_{ub}\right|$ & {$0.0037 \pm 0.0009$} \\
$\left| V_{cs}\right|$ & {$0.9771 \pm 0.0017$} \\
$\left| V_{cb}\right|$ & {$0.035 \pm 0.007$} \\
$\left| V_{td}\right|$ & {$0.007 \pm 0.002$} \\
$\left| V_{ts}\right|$ & {$0.035 \pm 0.007$} \\
$\left| V_{tb}\right|$ & {$0.9993 \pm 0.0002$}  \\
$m_{u}(2~{\rm GeV})_{\overline{\rm MS}}$ & $2.1 \pm 0.9$~{\rm MeV}   \\
$m_{d}(2~{\rm GeV})_{\overline{\rm MS}}$ & $4.2 \pm 1.4$~{\rm MeV}   \\
$m_{s}(2~{\rm GeV})_{\overline{\rm MS}}$ & $84 \pm 19$~{\rm MeV}  \\
$m_{e}^{\rm pole}$ & {$0.49 \pm 0.13 ~{\rm MeV}$} \\
$m_{\mu}^{\rm pole}$ & {$92 \pm 17 ~{\rm MeV}$} \\
\hline \hline
\end{tabular} 
\caption{\rm  Low energy measured values of 
$\left| V_{us}\right|$, the ratio of running masses $m_{c}/m_{t}$ and the 
$1~\sigma$ global fit of the phase $\gamma$ 
are used to determine $\lambda$ and 
$\theta$ from Eqs.~\ref{usecmara} \&  \ref{vus}.
Then these are used to predict the three lighter quark masses, $\sin(2\beta)$,
the rest of CKM matrix elements and the muon and electron masses from
the Yukawa matrices in Eqs.~\ref{mUrecons1a} , \ref{mdCPgamma} and \ref{mlmat}.}
\label{predictions}    
\end{table} 
while $\beta$ and $\gamma$ are related to the next order in $\lambda$ by,
\begin{equation}
\beta =  {\rm Arg} \left[ (2 - e^{-i \gamma}) \left( 1 + \theta \lambda (1-  2e^{i \gamma}) \right)
\right].
\label{betagamma2}
\end{equation}
We use as an input 
three parameters determined experimentally:
$\left| V_{us} \right|$, $m_{c}/m_{t}$ and the angle $\gamma$,
whose numerical values can be read in table~\ref{predictions}.
We determine $\lambda$, $\theta$ solving numerically
the system of Eqs.~\ref{usecmara} and \ref{vus}. 
Finally we use these together with the measured 
third generation fermion masses to predict
the masses of the lighter quarks, the rest of the CKM elements
and $\sin(2\beta)$,
the results are shown in table~\ref{predictions}. 
It is remarkable that 
all the next to leading order predictions 
agree with the respective measured values.
We note that the value predicted for $\sin(2\beta)$ is 
slightly lower than the value predicted to leading order in the previous
section.

It is worth to note that a theory of flavor must predict succesfully the
so-called $Q$ factor. This is a combination of quark masses which
has been determined experimentally 
from pseudoscalar meson masses to a $3.5$~\% accuracy.
 It is defined by,
\begin{equation}
Q = \frac{\frac{m_{s}}{m_{d}}}{\sqrt{1- \left( \frac{m_{u}}{m_{d}}\right)^{2}}} = 22.7\pm 0.08.
\end{equation}
In our case using the central values for $\theta$, $m_{t}$, $m_{b}$ and $\gamma$
in table~\ref{predictions} and $\lambda=0.211 \pm 0.007$ we obtain
$Q= 23.5 \pm 0.80$ which agrees at 1$\sigma$ with the experimental result.
For $\lambda= 0.218$ we obtain the central value $Q=22.7$.
We note that it is not convenient to use the measured value of Q to 
determine one of the basic parameters, instead of $\left|V_{us}\right|$ 
or $m_{c}/m_{t}$, because Q contains a implicit
dependence on the uncertainity in the top and bottom quark masses.

To sum up, there are six input parameters in the up, down and charged lepton Yukawa matrices:  
$\theta$, $\lambda$, $\gamma$, $m_{b}(m_{b})$, $m_{t}^{\rm pole}$ and $m_{\tau}^{\rm pole}$. 
We can choose two observables to determine $\theta$ and $\lambda$. In table~\ref{predictions} 
we chose $\left|V_{us}\right|$ and the charm/top quark mass ratio. Therefore we obtain the following true predictions: two quark mixing angles, a prediction for the CP-phase $\beta$, the up, down and strange quark masses plus the electron and muon masses. A total of 8 true succesfull predictions as can be seen in table II. 
\section{Charged lepton sector spectra and the Georgi-Jarlskog factor
\label{leptonsec}} 
We pointed out in Sec.~\ref{correlations},
see Eqs.~\ref{lambdaEq} and \ref{thetaEq},
that there are empirical relations that connect 
the charged lepton and the quark masses. 
In this section we argue that there is already a 
simple explanation for these relations, the Georgi-Jarlskog factor.
Let us assume that the normalized Yukawa matrix for the charged 
lepton sector is given by, 
\begin{equation}
\widehat{\bf m}_{L} = 
\left(
 \begin{array}{ccc}
 0 & \theta \lambda^{2} &e^{-i\gamma} \theta \lambda^{2}  \\
\theta \lambda^{2} &  
{\bf 3}~ \theta \lambda &   2 \theta \lambda \\
 e^{i\gamma} \theta \lambda^{2}  & 2 \theta \lambda & 1
\end{array}
\right).
\label{mlmat}
\end{equation}
If this were the case the charged lepton mass ratios 
could be calculated by a biunitary diagonalization,
\begin{eqnarray}
\frac{m_{e}}{m_{\mu}} &=& \frac{1}{9} \lambda^{2} ( 1 - 4 \theta \lambda ( \cos \gamma - \frac{17}{12} ) ), 
\label{leptonemu}\\
\frac{m_{\mu}}{m_{\tau}} &=& 3 \theta \lambda ( 1 - \frac{4}{3} \theta \lambda + \frac{1}{9} \lambda^{2}). 
\label{leptonmuta}
\end{eqnarray}
which would explain the observed empirical formulas. 
A matrix like Eq.~\ref{mlmat}, especially the
relation $(\widehat{\bf m}_{L})_{22} = 3 (\widehat{\bf m}_{D})_{22}$,
could be understood in the context of grand unified models. 
For instance, by embedding the quark and lepton fields in the 
representations ${\bf 5}$ and ${\bf 10}$ of SU(5) and  
assuming a non-minimal Higgs structure in the unified theory
\cite{Georgi:1979df} such that the field that couples to the matter fields
generating the $(\widehat{\bf m}_{L})_{22}$ entry
transforms under the representation ${\bf 45}$ of SU(5).
The corresponding Clebsch-Gordan factors could generate the 
factor 3 in the entry (22) of the charged lepton Yukawa matrix.
We must emphasize that even tough the original GUT model by Georgi and Jarslkog 
is ruled out the Georgi-Jarslkog factor, or in other words the ${\bf 45}$ Higgs,
has been used by many models especially supersymmetric GUT models which are not ruled out
by current data. This may indicate that the empirical relations support a mechanism which can be implemented in many GUT models but it does not support a particular GUT model.

Using the measured charged lepton mass ratios and
the 1$\sigma$ global fit value of $\gamma$ 
we can determine the parameters $\lambda$ and $\theta$
in the charged lepton sector from Eqs.~\ref{leptonemu} and \ref{leptonmuta}. 
Then using the top and bottom quark masses and the results of the previous section
we can predict the four lighter quark masses, the CKM elements and $\sin(2\beta)$.
The results are presented in table~\ref{leptonpredictions}.
All the predictions are very close to the respective measured values,
which is consistent with the numerical results in the previous section.

Alternatively one can use the values of $\lambda$ and $\theta$ as determined
from the quark sector together with the measured tau lepton mass 
to predict the electron and muon masses from Eqs.~\ref{leptonemu} and \ref{leptonmuta}.
The results, which are shown in table~\ref{predictions}, are consistent with
the measured electron and muon physical masses. 
\begin{table}
\begin{tabular}{|c|c|}
\hline 
\hline 
\multicolumn{2}{|c|}{input parameters } \\ 
\hline
$m_{e}/m_{\mu}$ &  $ (4.73711 \pm  0.00007 ) \times 10^{-3}$ \\
$m_{\mu}/m_{\tau}$ & $(5.882 \pm 0.001 ) \times 10^{-2}$ \\
$\gamma$  &  $61^{\circ}\pm 11^{\circ}$ \\
$m_{t}^{\rm pole}$ &  { $174.3 \pm 5.1 ~{\rm GeV}$} \\
$m_{b}(m_{b})_{\overline{\rm MS}}$ & {$4.2 \pm 0.1 ~{\rm GeV}$} \\
\hline
\hline 
\multicolumn{2}{|c|}{predictions} \\ 
\hline
$\lambda$ & $0.199 \pm 0.001$  \\
$\theta$ &   $0.100 \pm 0.001$ \\
$\sin(2 \beta)$  &   $0.820\pm 0.005$ \\
$\left| V_{ud}\right|$ & {$0.9778 \pm 0.0005$} \\
$\left| V_{us}\right|$ & {$0.211 \pm 0.003$} \\
$\left| V_{ub}\right|$ & {$0.0040 \pm 0.0001$} \\
$\left| V_{cd}\right|$ & {$0.211 \pm 0.003$} \\
$\left| V_{cs}\right|$ & {$0.9794 \pm 0.0002$} \\
$\left| V_{cb}\right|$ & {$0.0405 \pm 0.0006$} \\
$\left| V_{td}\right|$ & {$0.0075 \pm 0.0009$} \\
$\left| V_{ts}\right|$ & {$0.0401 \pm 0.0007$} \\
$\left| V_{tb}\right|$ & {$0.99917 \pm 0.00002$}  \\
$m_{c}(m_{c})_{\overline{\rm MS}}$ & $1.34 \pm 0.08$~{\rm GeV} \\
$m_{u}(2~{\rm GeV})_{\overline{\rm MS}}$ & $1.85 \pm 0.17$~{\rm MeV}  \\
$m_{d}(2~{\rm GeV})_{\overline{\rm MS}}$ &  $4.2 \pm 0.3$~{\rm MeV} \\
$m_{s}(2~{\rm GeV})_{\overline{\rm MS}}$ &  $94 \pm 5$~{\rm MeV} \\
\hline \hline
\end{tabular} 
\caption{\rm 
Low energy dimensionless ratios of measured charged lepton running masses plus 
the $1~\sigma$ global fit value of the phase $\gamma$ are used to determine $\lambda$ and $\theta$ from 
Eqs.~\ref{leptonemu} and \ref{leptonmuta}.
Then these are used to predict the four 
lighter quark masses, $\sin(2\beta)$ and the CKM 
matrix elements from Eqs.~\ref{mUrecons1a}, \ref{mdCPgamma}
and \ref{betagamma2}.}
\label{leptonpredictions}    
\end{table} 
\section{Perspectives and conclusions
\label{thetatheo}} 
In this section 
we include some considerations regarding the
possible characteristics of an underlying theory of flavor 
that is able to make sense of the previous results. 
To recapitulate,
\begin{enumerate}
\item 
There are two empirical formulas that connect the six fermion
mass ratios and the CKM elements,
\item
A simple three-parametric set of Yukawa matrices for the quark and
charged lepton sectors can generate these relations naturally
and account for the measured fermion mass ratios 
and CKM elements,
\item
The simplest explanation of the charged lepton hierarchies requires
the use of grand unification to account for the factor 3 in entry (22)
of the charged lepton Yukawa matrix,
\item
The proposed empirical formulas work perfectly at low energies 
but seem to get spoiled when extrapolated to very high energies, 
of the order of the GUT scale $M_{G} \approx 10^{16}$~GeV.
\end{enumerate}
Additionally, the proposed Yukawa matrices have the following characteristics,
\begin{enumerate}
\item
All the entries 
except entry (33) are proportional to a 
common parameter, $\theta$, which is approximately 
$\theta \approx0.1$
\item 
The generation of the correct fermion mass hierarchies requires the 
introduction of different powers of $\lambda$, the second flavor parameter,
which is approximately $\lambda\approx 0.21$
\item 
The CP violating phases $\beta$ and $\gamma$ are related by a simple
formula, which predicts that the maximum value of 
$\beta$ that can be reached is close to the measured value
\end{enumerate}
We note that any theory of flavor that can
generate the simple set of matrices proposed in this paper
(or an alternative set of matrices equivalent to leading order in $\lambda$) 
would automatically fit the experimental data.
The generation of hierarchies in the Yukawa matrices, like 
the hierarchies generated by polynomial matrices 
in powers of $\lambda$, is relatively easy to implement
through the breaking of a flavor symmetry, assuming that 
the vevs of the flavor breaking fields have a certain 
hierarchical structure. 
There are two characteristics I want to highlight:  
the presence of a common 
parameter in all the entries of the Yukawa matrices except entry (33),
and the fact that the empirical relations seem to get spoiled 
when extrapolated to very high energies.
A possible theory to explain the presence of a common 
parameter in all the entries of the Yukawa matrices except entry (33)
is the radiative generation of Yukawa couplings.
We note that the parameter $\theta$ has the right size to be 
a loop factor. If this is so, the Yukawa couplings must be generated
at a scale not very high; otherwise our mass relations would be spoiled,
as was pointed out above. To generate Yukawa couplings radiatively, one
has to postulate the existence of additional fields belonging to two different
sectors: the flavor breaking sector and the flavor messenger sector.
The messenger sector fields would transmit flavor violation from the
flavor breaking sector to the matter sector, generating Yukawa couplings radiatively. 
 One more piece of the puzzle is 
the factor 3 in the charged lepton Yukawa matrix;
the simplest explanation of this factor requires grand unification. 

Interestingly, supersymmetric models
can reconcile the generation of the Yukawa 
couplings at low energy with grand unification \cite{Ferrandis:2004ri}. 
It is known that grand
unification in the context of supersymmetric models can successfully predict
the weak mixing angle if the unification
scale is around $10^{16}$~GeV. This provides us with a 
consistent scenario where we can generate the Georgi-Jarlskog factor.
On the other hand, the presence of soft supersymmetry breaking terms allows 
for the radiative generation of quark and charged lepton masses through 
sfermion--gaugino loops.
The gaugino mass provides the violation of fermionic chirality required by a fermion
mass, while the soft breaking terms provide the violation of 
chiral flavor symmetry \cite{wyler}. In this case the superpartners of the
Standard Model matter fields would be the flavor messengers.
A supersymmetric model that implements the low energy radiative 
generation of Yukawa couplings has been proposed recently.
This was achieved by postulating a U(2) horizontal symmetry 
\cite{Barbieri:1995uv} that is broken by a set of supersymmetry breaking fields
\cite{Ferrandis:2004ri}. The model can also
overcome the present constraints on supersymmetric 
contributions to flavor changing processes \cite{Ferrandis:2004ng}.
\newline

It is known that 13 out of the 18 parameters of the Standard Model
belong to the flavor sector: 9 fermion masses, 3 mixing angles and
1 CP-violating phase. We have shown in this paper that there are regularities
underlying the measured fermion masses that allow us to connect them
through two simple empirical formulas. This implies a reduction in the
number of fundamental parameters in the underlying theory of flavor
from 13 to 6. We have proposed a simple set of three-parameter Yukawa
matrices, with two real parameters and a complex phase, 
that can precisely account for these mass relations and give us
a simpler parametrization of the CKM matrix 
The proposed set of Yukawa matrices may make the features
of the underlying theory of flavor more apparent.
Any theory of flavor that is
able to generate this set of matrices would automatically 
fit the experimental data. Furthermore, the proposed Yukawa matrices predict
a simple and succesfull relation between the SM CP-violating phases. We have also pointed out
that the empirical mass formulas between the quark and charged lepton masses 
find their simplest explanation in the context of grand unified theories.
There is hope that our knowledge of the ligther quark masses is going
to improve considerably in the near future by the use of lattice QCD methods.
These empirical relations, if confirmed, could be a guiding light in the search
for the underlying theory of flavor. 
\section*{Appendix\label{appendix}}
\subsection{Running Lepton masses \label{runleps}}
To compute the running charged lepton masses, I use well known
expressions, included here for completeness.
The physical charged lepton masses are related to the $\overline{\rm MS}$
running lepton masses, $m_{l}(\mu)_{\overline{\rm MS}} = m_{l}(\mu)$,
 through the relation,
\begin{equation}
m_{l}(\mu)= m_{l}^{\rm pole} \left( 
1 + \Delta_{l} + \Delta_{Z} +\Delta_{W} \right)
\end{equation}
where the one--loop self--energy correction is given by,
\begin{equation}
\Delta_{l}=  \frac{\alpha (\mu)}{\pi} \left[  \frac{3}{2}  \ln \left( \frac{m_{l}(\mu)}{\mu} \right) -1 \right],
\end{equation}
and the $Z$ and $W$ boson thresholds are given by,
\begin{eqnarray}
\Delta_{Z}&=& \frac{\alpha (\mu)}{4 \pi c^{2}_{W}} \left[
\left( 3 - 6 s^{2}_{W} + \frac{1}{4 s^{2}_{W}} \right)  
\ln \left( \frac{\mu}{m_{Z}} \right) \right. \nonumber \\
&&\left.+  \frac{7}{4} \left(  1 - 2 s^{2}_{W} + \frac{1}{28 s^{2}_{W}} \right)  \right], \\
\Delta_{W} &=&  \frac{\alpha (\mu)}{8 \pi } \left[
 \ln \left( \frac{\mu}{m_{W}} \right)  + \frac{1}{4} \right].
\end{eqnarray}
Using the measured physical masses,
\begin{eqnarray}
m_{e}^{\rm pole} &=& 0.510998902 \pm 0.000000021 ~\hbox{MeV}, \\
m_{\mu}^{\rm pole} &=& 105.6583568 \pm 0.0000052~\hbox{MeV},  \\
m_{\tau}^{\rm pole} &=& 1776.99\pm 0.3~\hbox{MeV}, 
\end{eqnarray}
we can calculate the running masses at a common scale.
We choose to evaluate the running masses at 
$\mu=m_{Z}$ where the self-energy correction dominates the threshold. 
We will use $s^{2}_{W} (m_{Z})_{\overline{\rm MS}}  =0.23113(15)$,  
$\alpha(m_{Z})^{-1}_{\overline{\rm MS}}  = 127.934 \pm 0.027$, 
$m_{W}= 80.423\pm 0.039$~GeV and $m_{Z}=91.1876\pm0.0021$~GeV. 
We obtain,
\begin{eqnarray}
m_{e}( m_{Z}) &=& 0.487304 \pm 0.000005 ~\hbox{MeV}, \\
m_{\mu} ( m_{Z})&=& 102.8695 \pm 0.0005~\hbox{MeV},  \\
m_{\tau}( m_{Z}) &=& 1748.87 \pm 0.30~\hbox{MeV}. 
\end{eqnarray}
We note that at $Q=m_{Z}$ 
the larger uncertainty in the running masses
comes from the uncertainty in $\alpha (m_{Z})$.
These running masses were used in Sec.~\ref{correlations} to search
for correlations in higher-order dimensionless ratios of charged
lepton masses.
\subsection{Running quark masses \label{runquarks}}
To calculate the dimensionless ratios of running quark masses we
must renormalize the quark masses to a common scale. 
For completeness we include in this section a brief explanation of the methods
used to calculate running quark masses and an update 
of previous numerical results \cite{quarkmasses}.
Different quark masses are usually given at 
different renormalization scales.  
For the top quark our starting point is 
the pole mass. We use the CDF/DO working group
average \cite{Hagiwara:fs}, 
\begin{eqnarray}
m_{t} &=& 174.3 \pm 5.1 ~\hbox{GeV}.
\label{toppole}
\end{eqnarray}
For the  bottom and charm quarks we start with 
the running masses, $m_{b}(m_{b})_{\overline{\rm MS}}$ 
and $m_{c}(m_{c})_{\overline{\rm MS}}$ as extracted from sum
rules in Refs.~\cite{bottommass} \& \cite{charmmass} respectively.
The averaged values are,
\begin{eqnarray}
m_{b}(m_{b})_{\overline{\rm MS}} &=& 4.2 \pm 0.1 ~\hbox{GeV},  
\label{mbmb} \\
m_{c}(m_{c})_{\overline{\rm MS}} &=& 1.28 \pm 0.09 ~\hbox{GeV}.
\label{mcmc}
\end{eqnarray}
This value of the charm quark mass is compatible with recent lattice calculations,
$m_{c}(m_{c})_{\overline{\rm MS}}^{\rm lat} = 1.26 \pm 0.16 ~{\rm GeV}$
~\cite{Becirevic:2001yh}.
For the lighter quarks we use
the normalized $\overline{\rm MS}$ values at $\mu=2$~GeV as extracted from sum rules
in Ref.~\cite{Jamin:2001zr,Gamiz:2002nu}. We use rescaled values \cite{Hagiwara:fs},
\begin{eqnarray}
m_{s}(2~\hbox{GeV})_{\overline{\rm MS}} &=& 117 \pm 17 ~\hbox{MeV},  \\
m_{d}(2~\hbox{GeV})_{\overline{\rm MS}} &=& 5.2 \pm 0.9~\hbox{MeV}, \\
m_{u}(2~\hbox{GeV})_{\overline{\rm MS}} &=& 2.9 \pm 0.6 ~\hbox{MeV}. 
\end{eqnarray}
We must add that there is a recent extraction of the strange 
quark mass by the HPQCD collaboration \cite{Aubin:2004ck}, 
using full lattice QCD, that has extracted a central value 
for the strange quark mass lighter that the one obtained 
by sum rules and has reduced considerably 
the corresponding uncertainity, 
$m_{s}(2{ \rm GeV})_{\overline{\rm MS}}^{\rm lat} = 76 \pm 10$ MeV.
This value has not been used in the main text because it has not yet 
been confirmed by other lattice QCD collaborations. 

For simplicity and to reduce the propagation of uncertainties
we rescale the top, bottom and charm quark masses 
down to $\mu=2$~GeV.
To calculate the running top quark mass at $\mu=2$~GeV we use 
the two-loop relation between the $\overline{\rm MS}$
and pole quark masses, which is known
through  order ${\cal O}(\alpha_s^3)$  
~\cite{Tarrach:1980up,Gray:1990yh,Fleischer:1998dw,Chetyrkin:1999qi,Chetyrkin:2000yt,Melnikov:2000qh},
\begin{eqnarray}
&& \frac{m (\mu) _{\overline{\rm MS}}  }{M} = 1
+a_s(\mu)
\left [L -\frac{4}{3}   \right ]
+a_s^2(\mu)
\left[  -\frac{3019}{288} 
\right.  \nonumber \\
&& \left. 
+ \frac{71}{144}n+
\left(\frac{445}{72} - \frac{13}{36} n \right)L
 + \left(-\frac{19}{24} + \frac{n}{12}\right) L^2 
\right.  \nonumber \\
&& \left. + \frac{\zeta_3}{6}  - \zeta_{2} \left( 
2+ \frac{2 }{3}\ln 2  - \frac{1}{3} n \right)
- \frac{\pi^2}{6}  \Delta      \right ],
\label{chetyrkininv}      
\end{eqnarray}
where $a_s(\mu) = \alpha_s(\mu)/\pi$ 
is the $\overline {\rm MS}$ strong coupling constant, 
$M$ is the on--shell mass, $L=\log(M^{2}/\mu^2)$, 
$n$ is the number of light quarks and $\Delta$,
$ \Delta  = \sum_{i\leq n} \left( \frac{m_i}{M} \right)$,
is a small correction due to light quark mass effects \cite{Gray:1990yh}. 

To calculate the charm and bottom quark running masses at
$\mu=2$~GeV we use 
the analytic solution of the renormalization      
group equation in the $\overline{\rm MS}$ scheme.
This was originally obtained       
at three loops ~\cite{Tarasov:au} and recently the four loop term      
has also been computed ~\cite{Chetyrkin:1997dh} and found to be very small.      
This takes the form,      
\begin{eqnarray}
\frac{ m (\mu) _{\overline{\rm MS}}}{\widehat{m}}
&=&  
\left( 2 \beta _0 a _s \left( \mu \right) \right)^{\gamma _0/\beta _0}
\left\{ 
1+\left( \frac{\gamma _1}{\beta _0} -\frac{\gamma _0\beta _1}{\beta _0^2}\right) 
a_s\left( \mu \right) + 
\right. 
\nonumber \\
&& + 
\left. 
\frac 12 
\left[ \left( \frac{\gamma _1}{\beta _0}- \frac{\gamma _0\beta _1}{\beta _0^2}\right) ^2+ \right. 
\right. 
\nonumber \\ 
&& 
\left. 
\left. 
\left( \frac{\gamma _2}{\beta _0}+\frac{\gamma _0\beta _1^2}{ \beta _0^3}-
\frac{\beta _1\gamma _1+\beta _2\gamma _0}{\beta _0^2} \right) 
\right] 
a_s\left( \mu \right)^2  
\right\}, 
\end{eqnarray}
where,
\begin{equation}
\begin{array}{l} 
\gamma _0 =1, \, \beta _0  =\left( 11-\frac 23n \right) \frac 14, \\  
\beta _1 = \left( 51- 
\frac{19}3n \right) \frac 18 ,\\ 
\gamma _1=\left(  
\frac{202}3-\frac{20}9n \right) \frac 1{16} ,\\ 
\beta _2=\left( 2857-\frac{5033}9n -\frac{%
325}{27}n^2\right) \frac 1{128} ,\\
\gamma _2 = \left( 1249-\left(  
\frac{2216}{27}+\frac{160}3\zeta (3)\right) n -\frac{140}{81}n^2\right) 
\frac 1{64}. 
\end{array}
\end{equation} 
Here $n$ is the number of light quarks, 
the integration constant $\hat {m}$ is     
the renormalization  group invariant mass. 
We do not need to know $\hat {m}$ because,      
if we denote the right hand side as $\hat {m}_{\rm MS} F_n(\mu)$, the running      
mass at scale $\mu$ can be calculated from a given running mass at scale $m(m)_{\overline{\rm MS}}$      
using the expression,      
\begin{eqnarray}      
m (\mu)_{\overline{\rm MS}} &=& m (m)_{\overline{\rm MS}} 
\frac{F_n(\mu)}{F_n(m(m))}.   
\label{mmumm}      
\end{eqnarray}      
In the case of four active light quarks we obtain,
\begin{eqnarray}      
F_{4} (\mu) = &&       
 \left(\frac{25 a_s(\mu)}{6} \right)^{12/25}      
\left(      
1 + \frac{3803}{3750}~a_s(\mu) 
\right.  \nonumber \\
&& \left. 
+ 2.078459~a_s^2(\mu)\right).      
\label{mcbrunms}      
\end{eqnarray}      
To compute these 
we need the values of $\alpha_s(m_b)$, $\alpha_s(m_c)$ and
$\alpha_s(\mu)$ 
corresponding to the experimental measurement at $\alpha_s(m_Z)$.      
To this end we use the three loop      
analytical formula for $\alpha_s$ in  the
$\overline{\rm MS}$ scheme~\cite{Tarasov:au}, which is the     
solution of the corresponding renormalization group
equation,
\begin{eqnarray}
\alpha _s(\mu )&=&\frac \pi {\beta _0 t }
\left[ 1-\frac{\beta _1}{\beta _0^2}\frac{\ln \left( t \right) }
{ t } \right. \nonumber \\
&&\left. +\frac{\beta _1^2}{\beta _0^4 t^2 }\left( \left( \ln \left(t \right) 
-\frac 12\right) ^2+\frac{\beta _2\beta _0}{\beta _1^2}%
-\frac 54\right) \right].   
\label{alfasrun}
\end{eqnarray}
Here $t=\ln 
\left(  \mu ^2 /\Lambda_{n} ^2\right)$ and $\Lambda_{n}$ is 
the integration constant for $n$ light quarks, to be determined from experiment.
The four loop contributions to Eq.~(\ref{alfasrun}) have      
also been calculated ~\cite{Vermaseren:1997fq} and found to be very small. 
In practice, we
use the value of $\alpha_s(M_Z)$ to first determine $\Lambda_{5}$
and $\alpha_s(m_b)$. Then we use $\alpha_s(m_b)$ to 
determine $\Lambda_{4}$, $\alpha_s(m_c)$  and $\alpha_s(\mu=2~\hbox{GeV})$.
Taking into account also the experimental uncertainty in $\alpha_{s}(m_{Z})$,
$\alpha_{s}(m_{Z})_{\overline{\rm MS}}=0.1172\pm0.0020$, we obtain, 
\begin{eqnarray}
\Lambda_{5}&=& 206 \pm 26~{\rm  MeV}, \\ 
\Lambda_{4}&=& 277 \pm 43~{\rm MeV}, \\
\alpha_{s}(m_{b}(m_{b}))_{\overline{\rm MS}} &=&  0.218 \pm 0.009,
\label{alphasb} \\
\alpha_{s}(2~\hbox{GeV})_{\overline{\rm MS}} &=&  0.286 \pm 0.019, 
\label{alphas2}  \\ 
\alpha_{s}(m_{c}(m_{c}))_{\overline{\rm MS}} &=&  0.354 \pm 0.043. 
\label{alphasc}
\end{eqnarray}
The uncertainty in $\alpha_{s}(m_{b}(m_{b}))$,
which is four times larger than the uncertainty in $\alpha_{s}(m_{Z})$, 
comes mainly from the uncertainty in 
the determination of $\Lambda_{5}$. The uncertainty in the determination
of $\Lambda_{5}$ and $\Lambda_{4}$ comes mainly from 
the uncertainties in $\alpha_{s}(m_{Z})$ and $\alpha_{s}(m_{b}(m_{b}))$
respectively.
Finally we can calculate the running quark masses at $\mu=2$~GeV.
We calculate the top quark mass from Eq.~\ref{chetyrkininv}
using as an input the top pole mass in Eq.~\ref{toppole} and
$\alpha_{s}(\mu)$ as determined in Eq.~\ref{alphas2}. We obtain, 
\begin{eqnarray}
m_{t}(2~\hbox{GeV})_{\overline{\rm MS}} &=&  298.2 \pm 15.4 ~\hbox{GeV}. 
\end{eqnarray}
The uncertainty comes from the top pole mass
uncertaninty and from the the uncertainty in the determination of $\alpha_{s}(\mu)$.
Alternatively one can compute the top quark running mass at the top scale,
$m_{t}(m_{t})$ using Eq.~\ref{chetyrkininv} and then use formula 
Eq.~\ref{mmumm} to calculate $m_{t}(\mu)$. These two approaches give
the same numerical results.
The charm and bottom quark running masses are calculated
from Eq.~\ref{mmumm}, using as an input the running masses,
$m_{b}(m_{b})_{\overline{\rm MS}}$ \& $m_{c}(m_{c})_{\overline{\rm MS}}$,
and the values of $\alpha_{s}(m_{b}(m_{b}))$, $\alpha_{s}(m_{c}(m_{c}))$
and $\alpha_{s}(\mu)$ determined in Eqs.~\ref{alphasb}--\ref{alphasc}.
We obtain,
\begin{eqnarray}
m_{c}(2~\hbox{GeV})_{\overline{\rm MS}} &=&  1.12 \pm 0.13 ~\hbox{GeV}  \\
m_{b}(2~\hbox{GeV})_{\overline{\rm MS}} &=&  4.91 \pm 0.20 ~\hbox{GeV}
\end{eqnarray}
Their respective uncertainties come mainly from the 
uncertainities in the theoretical extractions of 
$m_{b}(m_{b})_{\overline{\rm MS}}$ and $m_{c}(m_{c})_{\overline{\rm MS}}$
in Eqs.~\ref{mbmb}--\ref{mcmc}.
These running masses were used together with the
charged lepton running masses in the Sec.~\ref{correlations}. 
\acknowledgements
I thank Sandip Pakvasa and Xerxes Tata for valuable comments.
I thank H.~Guler for many suggestions.
This work is supported by the DOE grant number DE-FG03-94ER40833.

\end{document}